\documentclass{article}

\usepackage{arxiv}

\usepackage[utf8]{inputenc} % allow utf-8 input
\usepackage[T1]{fontenc}    % use 8-bit T1 fonts
\usepackage{url}            % simple URL typesetting
\usepackage{booktabs}       % professional-quality tables
\usepackage{amsfonts}       % blackboard math symbols
\usepackage{nicefrac}       % compact symbols for 1/2, etc.
\usepackage{microtype}      % microtypography
\usepackage{lipsum}		% Can be removed after putting your text content
\usepackage{graphicx}
\usepackage{natbib}
\usepackage{doi}

\usepackage{amsmath}
\usepackage{amssymb}
\usepackage{algorithm}
\usepackage[noend]{algpseudocode}

\usepackage[hang,small,bf]{caption}
\usepackage[subrefformat=parens]{subcaption}
\def\totalvardist{d_{TV}}
\def\kldist{d_{KL}}
\def\helldist{d_{H}}
\def\chidist{d_{\chi^2}}
\def\wassdist{d_{W_p}}
\def\domain{\mathcal{D}}
\def\M{\mathcal{M}}

\newcommand{\bcal}[1]{\mathbf{\mathcal{#1}}}

\DeclareMathOperator*{\argmin}{\bf{argmin}} % no space, limits underneath in displays
\title{Data-Driven Framework for Uncovering Hidden Control Strategies in Evolutionary Analysis}

\author{ \href{https://orcid.org/0000-0002-1543-5440}{\includegraphics[scale=0.06]{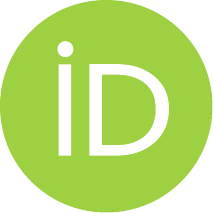}\hspace{1mm}Nourddine Azzaoui}\\
        University of Clermont Auvergne\\
        Clermont-Ferrand, France \\
	\texttt{nourddine.azzaoui@uca.fr} \\
	\And
	\href{https://orcid.org/0000-0003-3201-6106}{\includegraphics[scale=0.06]{orcid.pdf}\hspace{1mm}Tomoko Matsui} \\
	Department of Statistical Modeling\\
	The Institute of Statistical Mathematics\\
	Tokyo, Japan\\
	\texttt{tmatsui@ism.ac.jp} \\
	\AND
         \href{https://orcid.org/0000-0002-9544-9542}
	{\includegraphics[scale=0.06]{orcid.pdf}\hspace{1mm}Daisuke Murakami} \\
	    The Institute of Statistical Mathematics \\
	Tokyo, Japan \\
	\texttt{dmuraka@ism.ac.jp} \\
}

\renewcommand{\shorttitle}{\textit{arXiv} Template}

\begin{document}
\maketitle
\begin{abstract}
We have devised a data-driven framework for uncovering hidden control strategies used by an evolutionary system described by an evolutionary probability distribution. This innovative framework enables deciphering of the concealed mechanisms that contribute to the progression or mitigation of such situations as the spread of COVID-19. Novel algorithms are used to estimate the optimal control in tandem with the parameters for evolution in general dynamical systems, thereby extending the concept of model predictive control. This is a significant departure from conventional control methods, which require knowledge of the system to manipulate its evolution and of the controller's strategy or parameters. We used a generalized additive model, supplemented by extensive statistical testing, to identify a set of predictor covariates closely linked to the control. Using real-world COVID-19 data, we successfully delineated the descriptive behaviors of the COVID-19 epidemics in five prefectures in Japan and nine countries. We compared these nine countries and grouped them on the basis of shared profiles, providing valuable insights into their pandemic responses. Our findings underscore the potential of our framework as a powerful tool for understanding and managing complex evolutionary processes. 
\end{abstract}

% Keywords
% \keywords{First keyword \and Second keyword \and More}
%(list three to ten pertinent keywords specific to the article, yet reasonably common within the subject discipline.)
%}

\section{Introduction}
The evolution of natural dynamical systems is a complex process where the dynamics are influenced by a multitude of factors, including epidemic evolution, environmental changes, and various exogenous random events. 
Traditional methods of studying evolutionary processes often involve creating simplified models that may not fully capture this complexity. When it comes to influencing or controlling this evolution, optimal control theory (OCT) is a powerful and flexible tool \cite{grune_exponential_2020,bussell_applying_2019,ros_evidence_2004}.
This can provide valuable insights into the likely paths of evolution and the factors that may affect it \cite{shen_development_2019}.
The aim of this work was to investigate an evolutionary process, such as epidemic evolution, from the perspective of OCT in order to uncover the hidden mechanisms that lead to deterioration or improvement of the situation. We assumed that the evolutionary process is described by a controlled system via a hidden mechanism to which we do not have access and which we cannot change directly; we can only observe its effects through measurements related to its evolution. We also assumed that this hidden control mechanism is affected by an infinite number of random parameters related to the behavior and characteristics of societies and populations. We started from a simple linear control model based on Kalman-type filtering. We aimed to jointly determine the parameters of spatial-temporal process and the active optimal control that led to the observed evolution. The determination of the appropriate parameters was based on the best proximity between the observed data and that generated by the simple model.

Given that the objects in evolution are probability distributions, various customized distance measures were employed to evaluate their proximity. The Hellinger distance was thus preferred due to its superior numerical performance, inherited from its link to the $\ell_2$-norm \cite{HellingerTV}.
The coronavirus disease 2019 (COVID-19) has been examined using the principles of OCT. For instance, Péni et al. \cite{peni2020nonlinear} suggested an optimal control strategy based on model predictive control (MPC) theory for a compartmental epidemic model that depicts the dynamics of COVID-19 virus transmission. Péni and Szederkényi \cite{peni2021convex} explored the application of the MPC approach to a nonlinear compartmental model. The COVID-19 outbreak in Germany was studied using the epidemiological SIDARTH model to determine the best feedback control approach using MPC \cite{kohler2021robust}. Carli et al. \cite{carli2020model} presented a unique feedback control mechanism using an epidemiological SIR (susceptible, infectious, and recovered or removed)-based model and the MPC approach to help policymakers effectively mitigate the effects of the COVID-19 pandemic in numerous regions, including Italy, where the pandemic had reached its peak. The SIR model was explored by Morato et al. \cite{morato2020optimal} in relation to MPC strategies to control COVID-19 infections in Brazil. All of these studies used MPC to simulate epidemiological models like the SIR model. However, using simulation to investigate the COVID-19 pandemic is challenging due to the simplistic assumptions and limitations of the models, such as their inability to represent treatments, their fixed parameters, and their uncertainty resulting from data quality concerns. We thus aimed to identify the variables affected the spread of COVID-19 by using a data-driven model and a simple controlled Kalman-type dynamical system.

Our goal in this work was to make a simple assumption about the dynamics of COVID-19 virus transmission and then utilize data to determine the parameters governing it. An extensive initial analysis of Japanese data was performed to understand the dynamics of the epidemic in diverse prefectures (Tokyo, Osaka, Hokkaido, Fukuoka, and Okinawa). Their differing characteristics were compared and linked via a generalized additive model (GAM) to different socio-demographic structures and the related policies that were adopted in each prefecture.

After validating our data-driven model on the Japanese data, we examined whether it can be applied to other countries or geographical regions with different population structures and dynamics. We did this by applying an MPC-like procedure, as we did for the Japan data, to simultaneously obtain the dynamics and control parameters that contributed to the observed evolution of the epidemic in each country. We then performed a comparative analysis between the countries based on their management strategies, such as for vaccinations and lockdowns.

\begin{itemize}
\item \underline{Descriptive behavior of epidemic by country:} Using the epidemic's fitted controlling parameters for each country or region, we attempted to clarify the relationship between the parameters and various socio-demographic indicators as well as the various policies that governments implemented to reduce the epidemic's spread.
\item \underline{Comparative analysis and clustering:} Using the contrasting characteristics and indicators of the various countries, we grouped them on the basis of the similarities of their profiles. We then looked at the key variables that led to countries being grouped together. We also conducted additional investigations for each group to determine which factors contributed most to the grouping.
\end{itemize}

Our contributions are summarized below:

\begin{itemize}
\item We propose using a hidden controller framework in which the optimal control and the parameters for evolution in general dynamical systems including the COVID-19 epidemic dynamical system are simultaneously estimated. This transforms the linear evolution scheme into a nonlinear problem. 
\item We present an efficient optimization algorithm from the practical point of view that is based on a classical optimization algorithm, e.g., fmincon in MATLAB. 
\item We present the results of a comparative analysis between prefectures and regions of Japan with differing population structures and epidemic management policies that were obtained using our hidden controller framework.
\item We show that generalizing the model using our hidden controller framework to other countries (Australia, Germany, Italy, Brazil, South Africa, and so on) can reveal the effect of different policies and population structures on the evolution of the epidemic. 
\end{itemize}

We discuss in detail the methodology we used as well as related work in section 2. In section 3, we describe our experiments along with the application of our proposed model to real-world datasets. In section 4, we discuss our results and the importance of our study. We conclude in section 5 with a summary of the key points.

\section{Methods}
\begin{figure}
\centering
\includegraphics[width =0.6 \textwidth]{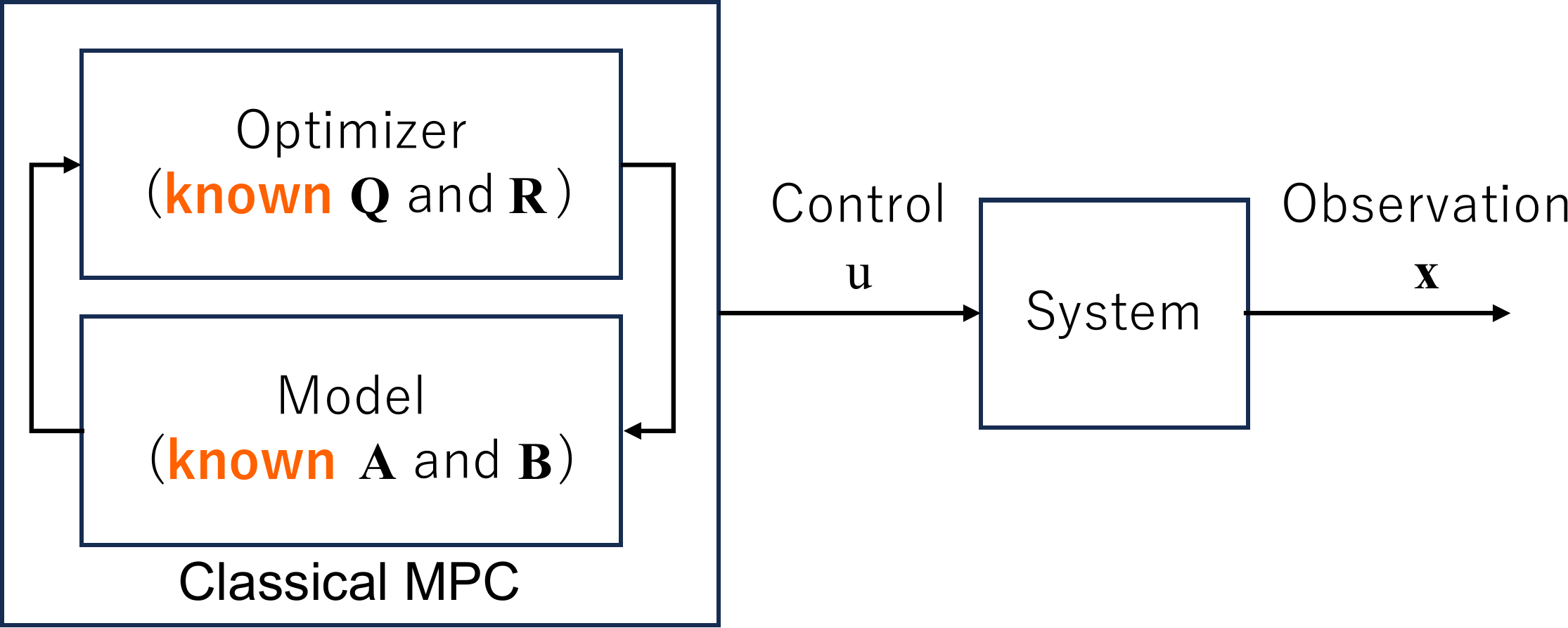}
\caption{In the MPC framework, users can access and manipulate (i.e., change or control) the system parameters.}
\label{fig:Scheme1}
\end{figure}

\begin{figure}
\centering
\includegraphics[width =0.6 \textwidth]{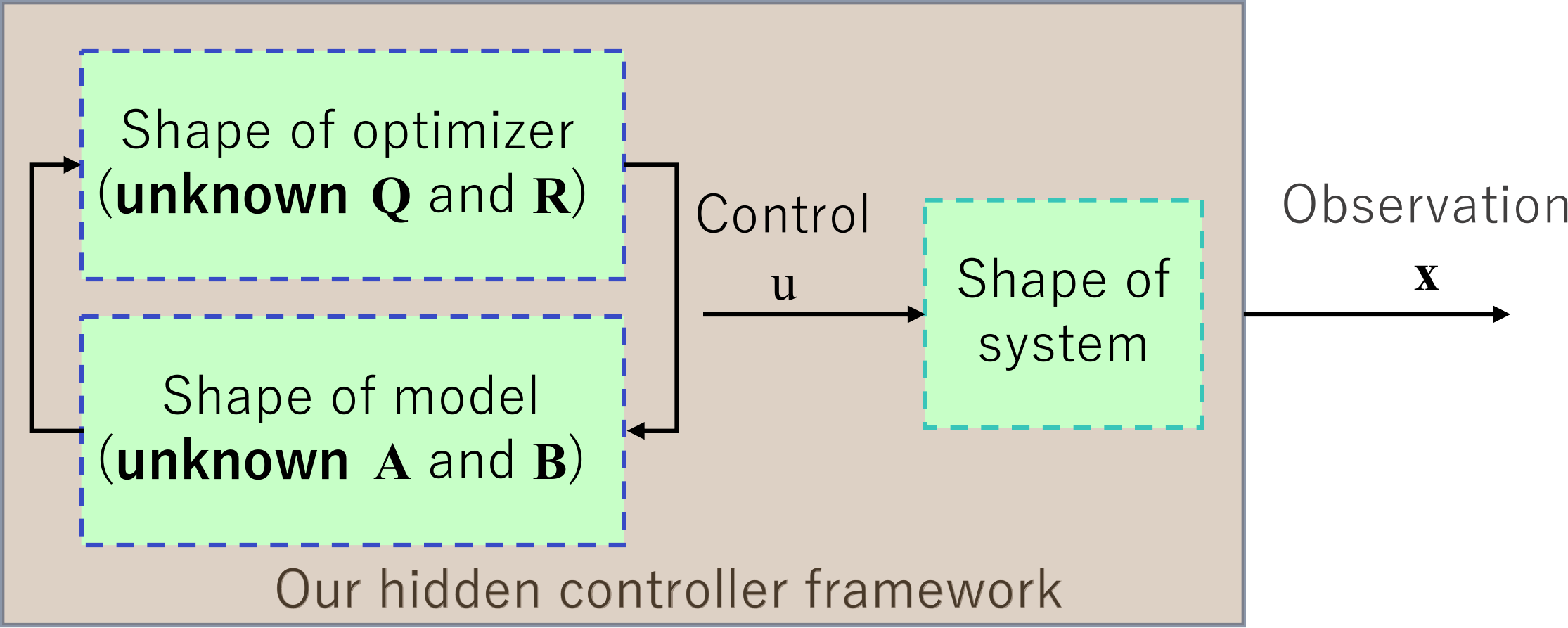}
\caption{In our framework, users do not have access to the items in the shaded area; they know only the shapes of the optimizer, model, and system and are unable to act on that information.}
\label{fig:Scheme2}
\end{figure}

The most commonly used data-driven control technique is based on MPC (Figure \ref{fig:Scheme1}). It enables the controller or user to interact with a dynamical system and affect its evolution by applying a specific control variable at selected times or epochs depending on the predicted horizon. Our approach was inspired by this technique but differs in two key aspects: we do not have direct access to the system, so we cannot manipulate its evolution, and we do not have any insight into the controller's strategy or parameters. (Here we focus on a Kalman-type dynamical system, but our approach can be applied to any parametric/non-parametric dynamical system). Our knowledge is limited to the fact that the controller or the system uses a model from a specific system family (here, for example, one in the form of equation (\ref{MMSLD0})). That is, we have access to only the observed final output data. This idea is sketched in Figure \ref{fig:Scheme2}.

This section is organized as follows; We begin in by reviewing related work and classical optimal control techniques in Sections \ref{presentation} and \ref{related}. In Sections \ref{Strategy} and \ref{interpret} we describe our proposed hidden control framework. 

\subsection{Targeted control problem and its formalism} 
\label{presentation}

We assume a discrete-time context in which decision-makers (e.g., government agencies) observe the distribution of the disease variable (e.g., the numbers of weekly infected, recovered and deceased individuals), possibly with observational inaccuracy, for a given time period (day, week, month, etc.). For each sub-population $P_i$, the observed distribution is generally estimated empirically from data that we denote $\widehat{\Pi}_i(k) = (\hat{\pi}^1_i(k), \hat{\pi}^2_i(k), \ldots, \hat{\pi}^{n+1}_i(k))$. The estimated evolutionary distributions for the whole population can be represented as
\begin{equation}\label{Mobs}
\mathcal{M}_k = \left(\begin{array}{ccc}
        \,\,\, \hat{\pi}^1_1(k) & \qquad \dots &  \qquad \hat{\pi}^1_p(k) \,\,\, \\
        \,\,\,\hat{\pi}^2_1(k) & \qquad \dots &  \qquad  \hat{\pi}^2_p(k) \,\,\, \\
       \,\,\, \vdots  &\qquad \dots & \qquad  \vdots  \,\,\, \\
        \,\,\,  \hat{\pi}^{n+1}_1(k) & \qquad \dots &  \qquad \hat{\pi}^{n+1}_p(k) \,\,\,
      \end{array} \right).
\end{equation}
\noindent Since $\pi^{n+1}_{i}(k)$ can be expressed as $\pi^{n+1}_{i}(k)= 1 - \displaystyle\sum_{\ell=1}^{n} {\pi^{\ell}_{i}(k)}$, we are interested in the dynamical evolution of the discrete $n$-dimensional probability vector $x_i(k)= (\pi^1_i(k), \pi^2_i(k) ,\ldots , \pi^n_i(k))$. Because we estimate the optimal control by sub-population, we omit the suffix $i$ hereafter to simplify the description (e.g., $x_i(k)$ is represented as $x_{k}$, and a column of $\M_k$ as $M_k$). This simplified description represents the state vector of the disease distribution at time $k$. To model the evolution, we propose using a discrete Kalman-like controlled system:
\begin{equation}
x_{k+1} = A \, x_{k} + B \, u_{k}, \ \forall \ k=1,\ldots,K-1,
\label{MMSLD0}
\end{equation}

\noindent where
\begin{itemize}
\item The term $u_{k} \in \mathbb{R}^{m}$ is an $m$-dimensional vector describing the unknown optimal control. This unknown control is the hidden mechanism that led to the observed shape of the evolution.\medskip
\item The deterministic state matrix $A =(a_{ij}) _{i,j=1,\dots, n } $, defined in $\mathbb{R}^{n\times n}$, and describing the logical evolution of $x_k$ when no control is applied is also unknown. \medskip
\item The unknown control matrix $B=(b_{ij})_{i=1,\dots, n, j=1,\dots,m}$, defined in $\mathbb{R}^{n\times m}$, describes how the control is applied to the system to bring it to its targeted state.
\end{itemize}
The idea is to use only observed data $\M_k$ in (\ref{Mobs}) to simultaneously estimate all the unknown objects of this control problem, namely matrices $A, B$ and the control term $u_k$ that leads to the observed shape of the data. When $A$ and $B$ are known, the optimal control $\hat{u}^{\ast}_k$ can be obtained using classical techniques that have been widely discussed in the literature (e.g., \cite{MCT, OAOC, LSTD}. In our case, these parameters are unknown and should be estimated from the observed data. Our strategy is to use the optimal control results obtained using an iterative algorithm that minimizes a cost function. 

\subsection{Control algorithm and related work}
\label{related}
For given deterministic matrices $A$ and $B$, the solution of the discrete-time optimal control problem (\ref{MMSLD0}) of a linear system is described in \cite{MCT, LSTD}. It consists of using a linear optimization scheme (\ref{MMSLD0}) and finding the optimal control $u= (u_1, \ldots, u_K)$ that minimizes the objective or cost function: 
\begin{equation}
J(u)= x^{'}_{K}\, S \, x_{K}+\sum^{K-1}_{k=1}{x^{'}_{k} \, Q\, x_{k}+ u^{'}_{k}\, R\, u_{k}},
\label{MPPRD}
\end{equation}
where exponent $(.)^{'}$ represents the transpose, and $Q, R$, and $S$ are cost matrices defined as follows.
\begin{itemize}
\item $Q \in \mathbb{R}^{n\times n}$ is a positive definite weighting matrix associated with the state vector. It is linked to the relative importance assigned to state vector $x_k$ in cost function $J(\cdot)$. In other words, a higher value in the diagonal of this matrix indicates a bigger penalty or greater emphasis placed on minimizing related state vector $x_k$.
\item $R \in \mathbb{R}^{m\times m}$ is a positive definite weighting matrix affecting the relative relevance or weight assigned to control $u_k$ in cost function $J(\cdot)$. It specifies how much emphasis the control algorithm should place on minimizing the control effort. A higher value in its diagonal, similar to that in the $Q$ matrix, indicates a higher penalty or greater importance placed on reducing the control input.
\item $S$ is a positive definite weighing matrix linked to the final state vector. It is a cross-weighting matrix used by cost function $J(\cdot)$ to penalize discrepancies between the inputs from state $x_k$ and control $u_k$. It is aimed at identifying any cross-dependencies between the states and control inputs. We are not interested in the final state here, so we suppose that $S=0$.
\end{itemize}
\noindent The optimal control problem has been widely studied (e.g., \cite{MCT, OAOC}). In our case, a linear quadratic regulator variant is used to minimize (\ref{MPPRD}). The solution of the state space equation (\ref{MMSLD0}) of a discrete time-invariant linear system has been given in a more general framework \cite{DCE} for $2 \leq k \leq K$:
\begin{equation}
x_{k} = A^{k}x_{1}+\sum_{i=1}^{K}{A^{k-i}Bu_{i}}.
\label{SEEE}
\end{equation}
This equation expresses state vector $x_{k}$ in terms of the initial state vector $x_{1}$ and a sequence of control vectors, $u_{i}, i=1,\ldots,k$, that are implicitly dependent: $A$, $B$ and weighting matrices $Q$, $R$. In this setting, the controlled system, (\ref{MMSLD0}), with dimension $K$ is controllable if the controllability matrix,
\begin{equation}\label{Controlability}
\mathbf{C}=\left[B, A B, \ldots, A^i B, \ldots, A^{K-1} B \right],
\end{equation}
has column rank $K$ (i.e., $K$ linearly independent columns). Under these conditions on parameters $A$, $B$, the optimal control solution \cite{DCE} is given by
\begin{equation}
u^{\ast}_{k} = - \bcal{K} \, x_{k}.
\label{oc}
\end{equation}
\noindent In this setting, the closed-form expression of the gain matrix $\bcal{K}$ is obtained using a Linear Quadratic Regulator (LQR). This is done iteratively in two steps using a backward approach, as detailed in Algorithm 1: %\ref{algorithm1}:

\begin{itemize}
\item Gain matrix $\bcal{K}$ is calculated from evolutionary parameter matrices $A$, $B$, $Q$, and $R$ via $S_{k+1}$ at step $k$:
\begin{equation}
\bcal{K}(A,B) = (R+B^{'}S_{k+1}B)^{-1}B^{'}S_{k+1}A.
\label{MDG}
\end{equation}

\item $S_k$ is obtained from the solution of the Riccati equation \cite{Ricatti}:
\begin{equation}
S_{k} = Q+A^{'}S_{k+1}A-A^{'} S_{k+1} B \bcal{K}(A,B). 
\label{S}
\end{equation}
\end{itemize}
\noindent By combining equations (\ref{MMSLD0}) and (\ref{oc}), we can write the evolutionary structure of the controlled system (\ref{MMSLD0}) recursively:
\begin{equation}
  x^*_{k}=(A - B \, \bcal{K}(A,B))x_{k-1}.
  \label{SEBF}
\end{equation}
In the general framework of optimal control, the parameters are known and fixed. In our study, we are dealing with unknown parameters. Therefore, our objective becomes the estimation of these unidentified matrices in the Kalman filter parameters, namely $A$, $B$, as well as weighting matrices $Q$, $R$. 

\subsection{Our hidden controller framework}
\label{Strategy}

\begin{figure}
\centering
\includegraphics[width=0.8\textwidth]{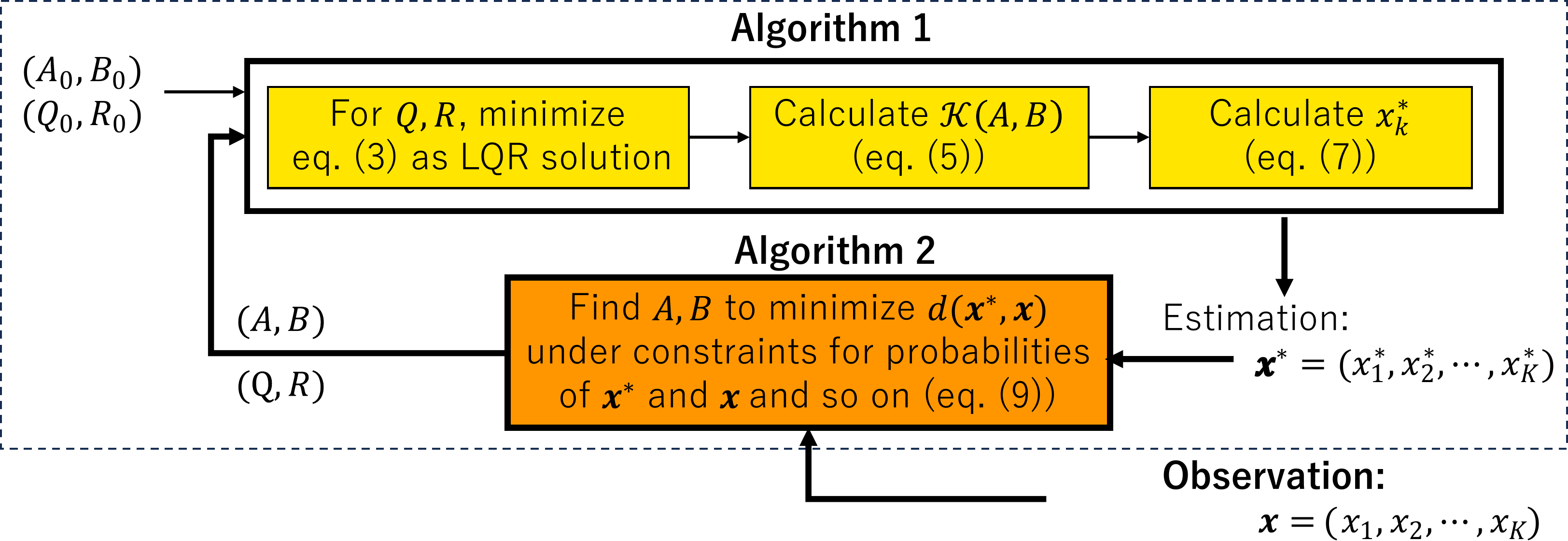}
\caption{Structure of main algorithm for estimating hidden control process}
\label{fig:Schema3}
\end{figure}

Instead of the traditional objects found in control problems, our methodology relies on using discrete probability distributions as the dynamic system object, as shown in (\ref{MMSLD0}), to characterize the system's behavior. A further distinguishing characteristic of our methodology is its ability to estimate both the parameters of the model and their corresponding optimal control, relying exclusively on observed data. To determine unknown parameters $A$ and $B$, our approach utilizes two interrelated optimization steps:
\begin{enumerate}
\item The first step is to find optimal control features $\bcal{K}(A,B))$ and $x^{\ast}_k$, as shown in (\ref{SEBF}) under controllability conditions on $A$ and $B$.
\item The second step is to minimize a \textbf{\emph{suitable distance}} that has been adapted to probability distributions between $x^{\ast}_k$ and observation $x_k$ under two constraints: $x_k$ is a discrete probability distribution calculated from observed data $M_k$ and the system is controllable. This leads to the following minimization problem:
\begin{align}
(\hat{A},\hat{B})=\argmin_{{\scriptscriptstyle{\substack{A, B , \text{s.t.} \forall t, \\ 0<\sum_t{(A-B\bcal{K})x_{t-1}}<1 }}}}
{\sum_{k=1}^{K}{d(x^{\ast}_{k},x_{k})} \label{PO}}.
\end{align}
The constrained minimum is taken over all $A$ and $B$ such that the system is controllable, as detailed in Algorithm 2. %\ref{algorithm2}.
\end{enumerate}
In accordance with the concept of the probability distribution, distance $d( . , . )$ in (\ref{PO}) determines the degree of similarity between two probability distributions. A variety of metrics and distances that have been used in different circumstances and scenarios have been reported. We next discuss five of them.

\subsection*{Metrics and distances} 

\begin{itemize}
\item The {\textbf{total variation distance}} measures the maximum difference between the probabilities that two distributions assign to an event. Therefore, it gives a \emph{worst-case} measure of the difference between the two distributions, which makes it a conservative discrepancy measure. It is given by 

\begin{align}\label{tv}
\totalvardist(D_1,D_2 )  &\triangleq \max_{S \subseteq \domain} (D_1(S)-D_2(S)) \nonumber \\
&= \frac{1}{2} \sum_{x \in \domain}|D_1(x)-D_2(x)| = \frac{1}{2} \|D_1 - D_2\|_{1}.
\end{align}
For the discrete case, the right-hand side of (\ref{tv}) can be expressed as the $\ell_1$ distance. Despite its frequent usage, the TV distance metric has certain disadvantages. It is sensitive to absolute differences between probabilities, which can lead to misleading results when applied to large event sets. Though it satisfies several metric properties like non-negativity, symmetry, and the triangle inequality, it is not a "proper" metric as it can assign a distance of zero to non-identical distributions. The TV distance also lacks the ability to capture dependencies among variables in multivariate distributions and does not provide information about the 'location' and 'spread' of differences. Furthermore, it is not a differentiable function, posing challenges for optimization algorithms. Lastly, its computational complexity increases substantially for continuous distributions, especially in high-dimensional scenarios. 

\item {\textbf{Kullback-Leibler (KL) divergence}} quantifies the information loss when distribution $D_1$ is used to approximate $D_2$:
  $$ 
 \kldist(D_1,D_2 ) = \sum_{\in \domain} D_1(x) \log\left(\frac{D_1(x)}{D_2(x)}\right).
 $$
KL divergence has several advantages and disadvantages when used as a measure of the difference between two probability distributions. Among its advantages are its high sensitivity to discrepancies in probability distributions, including those occurring in distribution tails, and its widespread application across multiple fields, such as information theory and machine learning. It also has a deep connection with maximum likelihood estimation, which enhances its value for theoretical analysis. The disadvantages include its lack of symmetry; that is, the divergence from distribution $D_1$ to $D_2$ is not the same as that from $D_2$ to $D_1$, which can make it counterintuitive as a "distance" measure. Furthermore, it is not a metric as it does not satisfy properties like symmetry and the triangle inequality. Another critical disadvantage is its undefined nature for non-overlapping distributions. Finally, when probability distributions are estimated from data, KL divergence can be excessively sensitive to the sample size, leading to potentially unreliable estimates for smaller sample sizes.

\item The {\textbf{Wasserstein distance}}, also known as the \emph{earth mover’s distance}, measures the minimum amount of work needed to transform one pile into another pile. For continuous or discrete probability distributions $D_1$ and $D_2$, the p-Wasserstein distance ($p\geq 1$) is defined as
  
  \begin{align*}
       \wassdist(D_1, D_2)  \triangleq & \inf \left\{\left(\mathbb{E}\left[ |X -  Y |^p\right]\right)^{\frac{1}{p}} ; \mathbb{P}_X = D_1 \text { and } \mathbb{P}_Y = D_2\right\}\\
    =  & \inf \left\{ ( \sum |x_i - y_i|^p P(X=x_i, Y=y_i) )^{\frac{1}{p}} \right\}.
  \end{align*}

The \emph{inf} bound is taken over all random variable couples $ (X, Y)$ such that the distribution of $X$ is $D_1$ and the distribution of $Y$ is $D_2$. It offers an intuitive geometric interpretation, acting as a measure of the minimum cost required to transform one distribution into another. In addition, it captures shifts in location, a property not captured by other metrics, making it advantageous in cases in which location is important. Moreover, unlike measures such as KL divergence, the Wasserstein distance remains defined even for non-overlapping distributions and exhibits robustness to noise. However, its computation can be resource-intensive due to the underlying optimization problem. It thus poses challenges, particularly in higher dimensions. The distance also depends on the choice of a cost function, which may not always have an evident best choice and can affect the distance calculation. Although sensitivity to location shifts can be beneficial, in instances where location is irrelevant, this sensitivity could capture unnecessary differences. Lastly, like many other measures, the Wasserstein metric faces scalability issues with high-dimensional data, demanding dimension reduction techniques to be effective.

\item The{\textbf{Hellinger distance}} measures the total difference between the square roots of the probabilities in the two distributions:
$$
d_H(D_1, D_2) = \frac{1}{2} \sum_{x \in \domain} \left(\sqrt{D_1(x)} - \sqrt{D_2(x)}\right)^2
 $$.
The Hellinger distance has several advantages. It possesses metric properties including symmetry and the satisfaction of the triangle inequality, making it a true metric. It is less sensitive to outliers due to the comparison of square roots of probabilities instead of the probabilities themselves. Its fixed range, typically between 0 and 1, aids in comparing distances across various distribution pairs. Moreover, its simplicity when comparing Gaussian distributions offers computational efficiency. However, the Hellinger distance is not without drawbacks. Its structure, rooted in square roots of probabilities, makes it less sensitive to differences in distribution tail behavior. It measures global dissimilarity, often failing to capture local differences effectively. Lastly, unlike KL divergence, it does not quantify information loss when one distribution approximates another, a property that may be desirable in certain contexts. 

\item The{\textbf{$\chi^2$ distance}} is the sum of the squared differences between the two distributions, divided by the first distribution. It is not symmetric.

$$
   \chidist(D_1, D_2) = \sum_{x\in \domain} \frac{(D_1(x) - D_2(x))^2}{D_1(x)}
$$
 \end{itemize} 

Each of these metrics comes with its own strengths and limitations. For instance, KL divergence ($\kldist$ ) and the chi-squared distance ($\chidist$) are non-symmetric, making them less intuitive for certain applications. On the other hand, metrics like the TV distance ($\totalvardist$) tend to be more conservative. In this study, our primary objective was to ensure robust convergence and computational efficiency in the optimization process. To identify the most suitable metric, we conducted comprehensive experiments testing these distances on our dataset. We encountered challenges regarding convergence and stability during the optimal control and estimation processes, as shown in (\ref{PO}). Our experiments revealed that the Hellinger distance ($\helldist$) exhibits superior performance in terms of convergence, stability, and efficiency. This can be attributed to its connection with the Euclidean $\ell_2$ norm, which provides certain computational advantages. Therefore, it is the primary distance measure we used in this study. 

\includegraphics{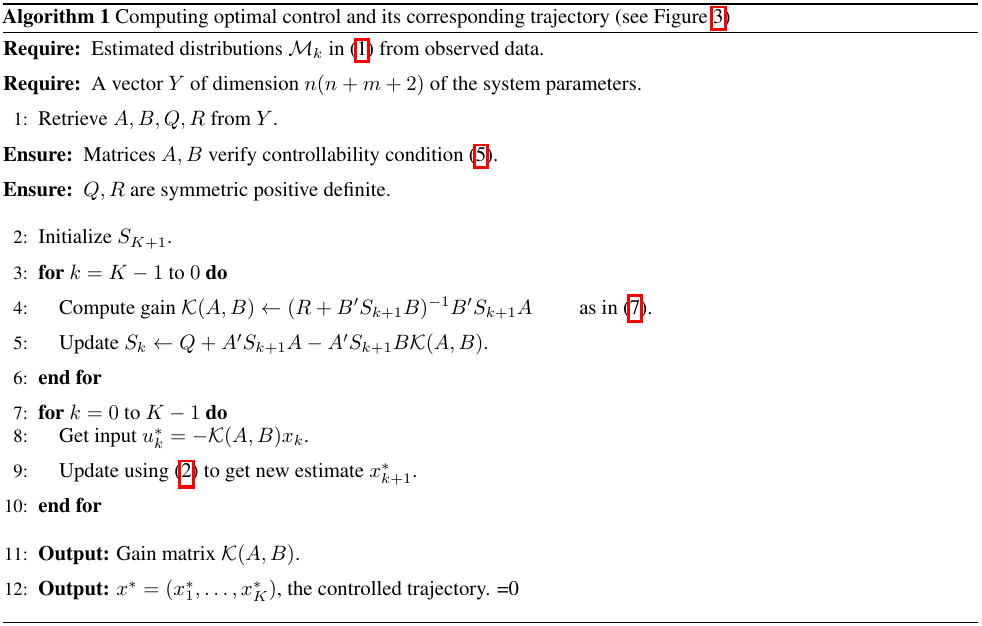} \label{algorithm1}
% \begin{algorithm}
% \caption{Computing optimal control and its corresponding trajectory (see Figure \ref{fig:Schema3})}
% \label{algorithm1}
% \begin{algorithmic}[1]
% \REQUIRE Estimated distributions $\mathcal{M}_k$ in (\ref{Mobs}) from observed data.\medskip
% \REQUIRE A vector $Y$ of dimension $n (n+m+2)$ of the system parameters.  \medskip
% \STATE Retrieve $A,B,Q,R$ from $Y$. \medskip
% \ENSURE Matrices $A,B$ verify controllability condition (\ref{Controlability}). \medskip
% \ENSURE $Q, R$ are symmetric positive definite.
% \bigskip
% \STATE Initialize $S_{K+1}$.\medskip

% \FOR{$k = K-1$ to $0$}\medskip

% \STATE Compute gain $ \mathcal{K}(A,B) \gets (R + B'S_{k+1}B)^{-1} B'S_{k+1} A$ \qquad as in (\ref{MDG}).\medskip

% \STATE Update $S_k \gets Q + A' S_{k+1} A - A' S_{k+1} B \mathcal{K}(A,B)$.\medskip

% \ENDFOR\medskip

% \FOR{$k = 0$ to $K-1$}

% \STATE Get input $u^{\ast}_k = -\mathcal{K}(A,B) x_k $. \medskip
% \STATE Update using (\ref{MMSLD0}) to get new estimate $x^{\ast}_{k+1}$.\medskip
% \ENDFOR

% \bigskip
% \STATE \textbf{Output:} Gain matrix $\mathcal{K}(A,B)$.\medskip
% \STATE \textbf{Output:} $x^{\ast} = (x^{\ast}_1, \ldots, x^{\ast}_K)$, the controlled trajectory. \medskip
% \end{algorithmic}
% \end{algorithm}

\subsection*{Our parameter estimation:} 
\label{estimation}
To prepare for the calculation of matrix parameters $A, B$ of the hidden closed-loop equation (\ref{SEBF}) on the basis of observed data, we vectorize them:  $(a_{11},\ldots, a_{nn},b_{11},\ldots, b_{nm})^{'} \in \mathbb{R}^{n^2 + nm}$. Furthermore, we assume that the hidden controller uses weighting matrices $Q$ and $R$ to calculate the gain matrix $\bcal{K}$ in (\ref{MDG}). There is no systematic way to make this choice, as it depends only on the user's preferences in terms of control over parameters $A$ and $B$. Most often, the weighting matrices $Q$ and $R$ are chosen to be diagonal. Due to the small number of observations, we use $Q= diag (\alpha_1, \dots, \alpha_n) $ and $R =diag(\beta_1,\dots,\beta_n)$. We then rearrange all the free parameters in a vector: $Y= (a_{11},\ldots, a_{nn},b_{11},\ldots, b_{nm}, \alpha_1, \dots, \alpha_n , \beta_1, \dots, \beta_n)^{'} \in \mathbb{R}^{n (n +m+2)} $.
Let us now consider the objective cost function: 
\begin{equation}\label{of}
F :  Y \longmapsto F(Y)=F(A,B, Q,R) =  \sum_{k=1}^{K}{d(x^{\ast}_{k},M_{k})},
\end{equation}
where $d(. , . )$ is one of the distances $\totalvardist$, $\helldist$, $\chidist$, $\wassdist$, or $\kldist$, and
\begin{itemize}
\item $M_{k}=\left( \hat{\pi}^1_i(k), \hat{\pi}^2_i(k), \ldots, \hat{\pi}^n_i(k) \right)^{'} \in \mathbb{R}^{n}$ is the statistically estimated distribution for population segment $i$ at observation time $k$. (Index $i$ is omitted for $M_k$ and $x_k$ to simplify the description, as mentioned in section \ref{presentation}.)
\item $x^{\ast}_{k}$ is the closed-loop state of the controlled system. The computation of vector $x^{\ast}_{k}$ for $k=1,\ldots,K$ is shown in (\ref{SEBF}). It is implicitly dependent on the parameters in vector $Y$. This step is nested inside the general optimization problem (\ref{PO}). Vector $Y$ should satisfy two constraints, $C_1$ and $C_2$, on the one hand, to comply with both the controllability constraints on $A$ and $B$ and the positive definiteness of $Q$ and $R$, and, on the other hand, to satisfy the probability structure of state vector $x_k$. 
\begin{enumerate}
\item[$\mathbf{C_1}$:] Vector $Y$ should verify the constraint that $A$ and $B$ lead to a controllable system by verifying (\ref{Controlability}).
\item[$\mathbf{C_2}$:] For all $k$, vector $x^{\ast}_k $ calculated using (\ref{SEBF}) should satisfy a probability structure, meaning that all its elements should be positive and their sum is less than 1. 
\end{enumerate}
\end{itemize}

The final optimal parameter vector, $\widehat{Y}$, is then obtained by solving the nonlinear optimization problem under constraints:
\begin{align}
   (\hat{A},\hat{B}, \hat{Q}, \hat{R}) = \widehat{Y} = \argmin_{{Y \in C_1\cap C_2  }}{\left[ F(Y) \right]}.
\end{align}

From the practical point of view and to reduce the relative error in estimating the unknown coefficients of the parameters, we assume that the hidden controller resets vector $x_{k}$ to the observed value each time when calculating it for interval $K$. To implement this assumption, we divide the range of time-ordered observations $x_{k}$ by a pre-selected value $T$ (e.g., for COVID dataset, $T=3$ (three days) or $T=7$ (one week)) and denote the number of observation groups formed by $L=[\frac{K}{T}]$, where $[x]$ is the integer part. We then consider the partition group $\left\{G_{1},\ldots,G_{L}\right\}$ on all the observations $\left\{M_{1},\ldots,M_{K}\right\}$ ordered in accordance with time: the observations in group $G_{1}$ being $G_{1}=\left\{M_{1},\ldots,M_{T}\right\}$, those in group $G_{2}$ being $G_{2}=\left\{M_{T+1}, \ldots,M_{2*T}\right\}$, and those in the last group being $G_{nstep}=\left\{M_{(L-1)*T+1},\ldots,M_{L*T}\right\}$, for example. For any $1\leq \ell \leq L$, we take vector $x_{k}$ as follows:
\begin{itemize}
\item $x_{k}=M_{k}$ for $k=(\ell-1)*T+1$. Thus, we reset vector $x_{k}$ to the first observed value in each group. 
\item $x_{k}=(A+B\bcal{K})*x_{k-1}$ for all $k=(\ell-1)*T+2,\ldots,T*\ell$.
\end{itemize}
For the solution of the nonlinear optimization problem (\ref{PO}), we use sequential quadratic programming, which is an efficient and widely used technique \cite{SQP, NO,PMOO,NLPASA}, Moreover, it is readily available in the MATLAB software used for the optimization part of this work.

\includegraphics{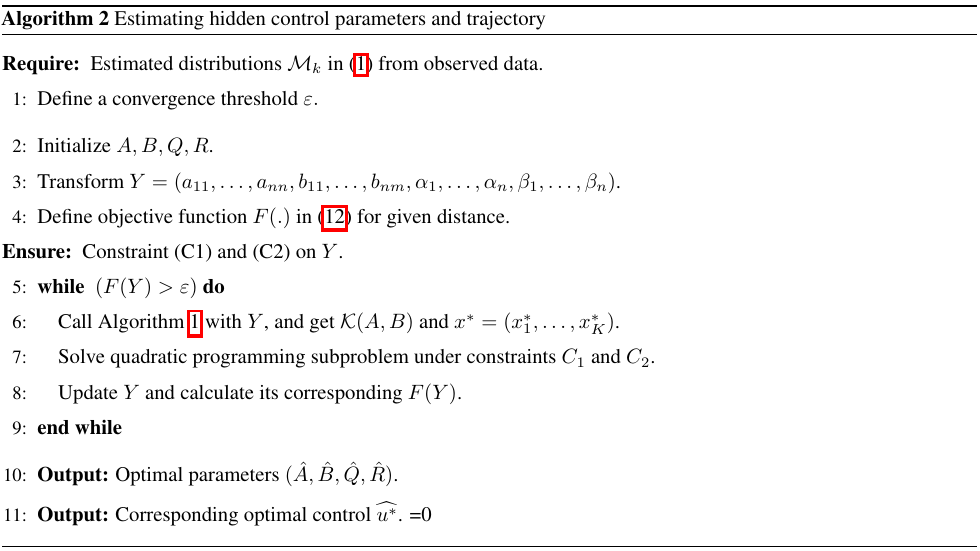} %\lavel{algorithm2}
% \begin{algorithm}[H]
% \caption{Estimating hidden control parameters and trajectory}\label{algorithm2}
% \medskip
% \begin{algorithmic}[1]
% \REQUIRE Estimated distributions $\mathcal{M}_k$ in (\ref{Mobs}) from observed data.\medskip
% \STATE Define a convergence threshold $\varepsilon.$\bigskip
% \STATE Initialize $A,B,Q,R$. \medskip
% \STATE Transform $Y= (a_{11},\ldots, a_{nn},b_{11},\ldots, b_{nm}, \alpha_1, \dots, \alpha_n , \beta_1, \dots, \beta_n)$.\medskip
% \STATE Define objective function $F(.)$ in (\ref{of}) for given distance.\medskip
% \ENSURE Constraint (C1) and (C2) on $Y$. \medskip

% \WHILE{ $\left( F(Y) > \varepsilon\right)$}\medskip
% \STATE Call Algorithm \ref{algorithm1} with $Y$, and get $\mathcal{K}(A,B) $ and $x^{\ast}=(x^{\ast}_{1}, \ldots, x^{\ast}_{K})$. \medskip
% \STATE Solve quadratic programming subproblem under constraints $C_1$ and $C_2$. \medskip
% \STATE Update $Y$ and calculate its corresponding $F(Y)$. \medskip
% \\

% \ENDWHILE
% \bigskip
% \STATE \textbf{Output:} Optimal parameters $(\hat{A}, \hat{B}, \hat{Q}, \hat{R})$.\medskip

% \STATE \textbf{Output:} Corresponding optimal control $\widehat{u^{\ast}}$.
% \medskip
% \end{algorithmic}
% \end{algorithm}

These algorithms were applied to each segment of the population. The trajectories estimated using these algorithms and the observed data for different countries are plotted in Figure \ref{pred-obse}.

\subsection{Interpretation of control parameters} 
\label{interpret}
In accordance with the preceding discussions, we computed the controlled part, denoted here as $\hat{B}, \widehat{u^{\ast}}$, of the resulting dynamical system. This computation enabled us to estimate the hidden parameters, and in turn, uncovered the controller's strategy. Another key accomplishment of this study was correlating these controlling elements with the characteristics of the corresponding population. The next step was to, a posteriori, explore how these elements affected the evolution of the dynamical system. To this end, we fitted the derived controlled part to the observed population characteristics as illustrated in Figure \ref{fig:Schema3}. For each population segment $P_i$, this was done using a multivariate GAM as such GAMs are well-known for their adaptability in dealing with various forms of data and their capacity to model nonlinear relationships. We conducted extensive statistical testing on the GAM output to identify the set of variables closely linked with the control part. 

\begin{figure}
\centering
\includegraphics[width=\textwidth]{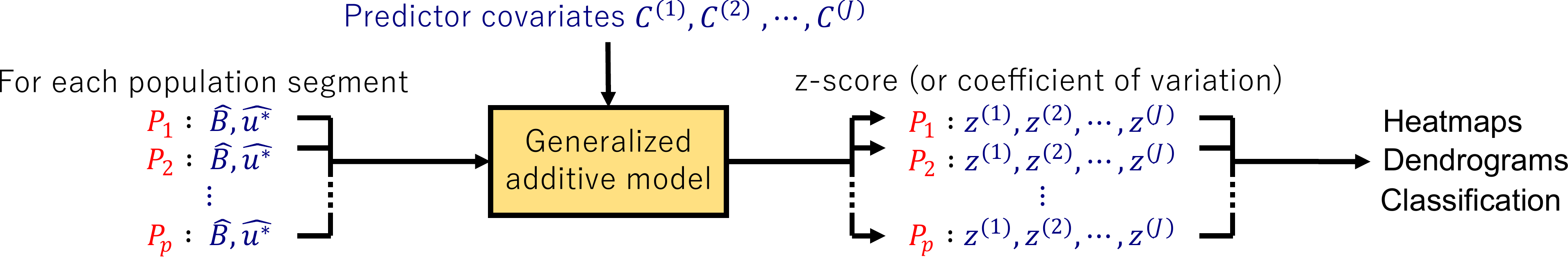}
\caption{For each population segment, the link between control terms $\widehat{B}, \widehat{u^{\ast}}$, and $C^{(1)}, \dots, C^{(J)}$, which describe the characteristic variables (covariates, predictors) of the population segment, was fitted to a GAM. The $z$-score output of the GAM was used for classification to identify how these predictors contributed to the control of the system.}
\label{fig:schema4}
\end{figure}

\paragraph{\textbf{Multivariate GAM}} Let us consider a multivariate GAM in which the response is vector $X=\hat{B} \, \widehat{u^{\ast}} \in \mathbb{R}^n $ and denote $C^{(1)}, \dots, C^{(J)}$ as the dependent variables describing the characteristics of the population. As shown in Table \ref{predictors}, we considered six characteristic COVID-19 variables specific to Japanese prefectures. The responses and covariates (predictors) are observed at times $k=1 \ldots K$. 

\begin{equation}\label{gam}
X = \mu_0 + S_1(C^{(1)}) + S_2(C^{(2)}) + \ldots +  S_J(C^{(J)}) + \varepsilon
\end{equation}
The functions $S_1(.), \ldots, S_J(.)$ are flexible smooth functions and can fit complex patterns in the data that would not be captured by a linear model. At the same time, they maintain interpretability: the fitted functions can be plotted and easily interpreted. The specifics of fitting a GAM depend on the type of response variable (whether it is binary, count data, continuous, etc.), the chosen link function, and the chosen form for the nonlinear functions $S_1(.), \ldots, S_J(.)$. This model has been well studied, and many software packages are available for fitting GAMs, such as Mixed GAM Computation Vehicle with Automatic Smoothness Estimation (mgcv) in R \cite{mgcv2011,mgcv2016}. As depicted in Figure \ref{fig:schema4}, the application of the GAM yields an array of newly minted comprehensible features designated as $z^{(1)}, \ldots , z^{(J)}$. These features serve to quantify the intensity of the association between the control and corresponding population characteristics, thereby offering a method to gauge the extent of a particular population characteristic's effect on the control of the obscured dynamical system.

These new features provide valuable information for investigating the dynamical system's behavior across various population segments. This facilitates the identification of similarities amongst segments for which the system exhibits identical behavior. Various classification techniques have been explored for achieving this functionality. More details, illustrative figures, and discussion can be found below on the experiments performed on actual COVID-19 data. 

\section{Experiments and application to real data}
\subsection{Data description}
We used two sets of observed COVID-19 data: (i) data from five prefectures in Japan provided by JX PRESS Corporation and (ii) data from nine countries sourced from a "worldwide epidemiological database for COVID-19" \cite{COVID19data} and the "COVID-19 data hub" \cite{ COVID19github}. Using our hidden controller framework, we estimated the unknown parameters $A, B$ and the gain matrix $K$ of the closed-loop system (\ref{SEBF}) to capture the dynamic behavior of COVID-19.

The prefectural data used in this study comprised the proportion of weekly infected individuals in prefecture $i$ ($\pi^1_i(k)$), the proportion of weekly recovered individuals ($\pi^2_i(k)$), the proportion of weekly deceased individuals ($\pi^3_i(k)$), and the proportion of all other people who never had contact with the virus ($\pi^4_i(k)) (=1 - \sum^3_{n=1}{\pi^n_i(k)})$. The proportions were estimated from weekly counts of numbers reported daily for 32 weeks from April 1, 2021, to November 10, 2021, coinciding with the vaccination campaign. The five prefectures ($i=1,\ldots,5$) comprise Tokyo, Osaka, Hokkaido, Fukuoka, and Okinawa, respectively. These prefectures are representative of Japanese populations for which there were a sufficient number of cases for statistical analysis. For each prefecture, the data are represented as a $32 \times 3$ matrix, with the weekly observations in the columns. The candidate weekly predictor covariates listed in Table \ref{predictors} are for the period two weeks prior. Factors such as increases in driving and public transport ("transit") in Tokyo, which are considered to affect the rate of new infections, are included. We used a two-week delay between measurement and onset in accordance with previous findings \cite{matsui_analysis_2022}.

\begin{table}%label{table1}
\begin{center}
\caption{Candidate weekly predictor covariates for five Japanese prefectures.}
\label{predictors} % Give a unique label
\begin{tabular}{|l|p{8 cm}|}
\hline\noalign{\smallskip}
Predictor covariate & Description \\ 
\noalign{\smallskip}\hline\hline\noalign{\smallskip}
\bf week & Time point (weekly ID) \\ 
\noalign{\smallskip}\hline\noalign{\smallskip}
\bf driving & Vehicle increase rate (provided by Apple Inc.; compared with January 13, 2020)\\ 
\noalign{\smallskip}\hline\noalign{\smallskip}
\bf transit & Public transport increase rate (provided by Apple Inc.; compared with January 13, 2020)\\ 
\noalign{\smallskip}\hline\noalign{\smallskip}
\bf walking & Pedestrian increase rate (provided by Apple Inc.; compared with January 13, 2020)\\ 
\noalign{\smallskip}\hline\noalign{\smallskip}
\bf num\_beds & Number of available beds in hospitals in Tokyo (provided by Ministry of Health, Labour and Welfare )\\ 
\noalign{\smallskip}\hline\noalign{\smallskip}
\bf vaccin & Number of people vaccinated in Tokyo (provided by Ministry of Health, Labour and Welfare )\\ 
\noalign{\smallskip}\hline
\end{tabular}
\end{center}
\end{table}

We selectively used data from nine countries in our experiments: Australia (AUS), Brazil (BRA), Chile (CHL), Colombia (COL), Czech Republic (CZE), Germany (DEU), Lithuania (LTU), South Africa (ZAF), and Japan (JPN). The proportion of weekly infected individuals in country $i$ ($\pi^1_i(k)$), the proportion of weekly recovered individuals ($\pi^2_i(k)$), and the proportion of weekly deceased individuals ($\pi^3_i(k)$) were calculated from the data on "confirmed" (cumulative number of confirmed cases), "deaths" (cumulative number of deaths), and "recovered" (cumulative number of patients released from hospitals or reported recovered)  \cite{COVID19data,COVID19github}. The period was the same as for dataset (i): the 32 weeks from April 1, 2021, to November 10, 2021.
The candidate predictor covariates for these nine countries are shown in Table \ref{predictors2}. Four of them represent governmental policy measures: government\_response\_index, stringency\_index, containment\_health\_index, and economic\_support\_index. Since these policy measures are correlated and the ranges of the index scores vary between countries, we first categorized the policy measures into three levels - low (L: 0–50 percentile), middle (M: 50–90 percentile), and high (H: 90–100 percentile), using the Japanese data as a reference. We then applied multiple correspondence analysis and hierarchical clustering on principal components to the categorized data and obtained three groups of policy measures, as shown in Table \ref{policy}. We again used a two-week delay between measurement and onset in accordance with previous findings \cite{matsui_analysis_2022}.

\begin{table}%\label{table2}
\begin{center}
\caption{Main categorical levels consisting of three groups of policy measures.}
\label{policy}
\begin{tabular}{|l|c|c|c|c|}
\hline\noalign{\smallskip}
index & government & stringency & containment & economic\\
 & \_response & & \_health & \_support \\
 \noalign{\smallskip}\hline\hline\noalign{\smallskip}
 \bf policy 1 & H & – & H & L \\
 \noalign{\smallskip}\hline\hline\noalign{\smallskip}
 \bf policy 2 & M & H & – & M \\
 \noalign{\smallskip}\hline\hline\noalign{\smallskip}
 \bf policy 3 & L & – & L & – \\
\noalign{\smallskip}\hline
\end{tabular}
\end{center}
\end{table}

\begin{table}%\label{table3}
\begin{center}
\caption{Candidate predictor covariates for nine countries \cite{COVID19data,COVID19github}.}
\label{predictors2} % Give a unique label
\begin{tabular}{|l|p{7.2 cm}|}
\hline\noalign{\smallskip}
Predictor covariate & Description \\ 
\noalign{\smallskip}\hline\hline\noalign{\smallskip}
\bf week & Time point (weekly ID) \\ 
\noalign{\smallskip}\hline\noalign{\smallskip}
\bf driving & Vehicle increase rate (provided by Apple Inc.; compared with January 13, 2020)\\ 
\noalign{\smallskip}\hline\noalign{\smallskip}
\bf transit & Public transport increase rate (provided by Apple Inc.; compared with January 13, 2020)\\ 
\noalign{\smallskip}\hline\noalign{\smallskip}
\bf walking & Pedestrian increase rate (provided by Apple Inc.; compared with January 13, 2020)\\ 
\noalign{\smallskip}\hline\noalign{\smallskip}
\bf hosp & Number of hospitalized patients\\ 
\noalign{\smallskip}\hline\noalign{\smallskip}
\bf people\_fully\_vaccinated & Number of people who received all doses prescribed by vaccination protocol \\
\noalign{\smallskip}\hline\noalign{\smallskip}
\bf government\_response\_index & Index of how government response varied over all indicators in Oxford COVID-19 Government Response Tracker database \cite{hale2021global}, becoming stronger or weaker over the course of the outbreak \\ 
\noalign{\smallskip}\hline\noalign{\smallskip}
\bf stringency\_index & Index of strictness of ‘lockdown-related policies that restricted people’s behavior\\ 
\noalign{\smallskip}\hline\noalign{\smallskip}
\bf containment\_health\_index & Index combining ‘lockdown’ restrictions and closures with measures such as testing policy, contact tracing, short term investment in healthcare, and investment in vaccines\\ 
\noalign{\smallskip}\hline\noalign{\smallskip}
\bf economic\_support\_index & Index representing such factors as income support and debt relief\\ 
\noalign{\smallskip}\hline
\end{tabular}
\end{center}
\end{table}

\subsection{Experimental conditions}
To calculate parameters $A, B$ of the closed-loop equation (\ref{SEBF}) on the basis of the observed COVID-19 data for the five Japanese prefectures (i) and nine countries (ii), we first defined $Y=(a_{11},a_{12},\ldots, a_{33},b_{11},b_{12},\ldots, b_{33})^{T} \in \mathbb{R}^{18=3*3 + 3*3}$ representing the unknown coefficients of parameters $A, B$. We then set the Hellinger distance (described in section \ref{Strategy}) to $d(x^{\ast}_{k},x_{k})$, where the term $x^{\ast}_{k}$ represents the estimated COVID-19 dynamics, and the other term $x_{k}=\left(\pi_{j}^{1}(k), \pi_{j}^{2}(k), \pi_{j}^{3}(k)\right)^{'} \in \mathbb{R}^{3}$ represents the observed COVID-19 dynamics on the $k^{th}$ day. 

The proportions $\Pi(k)$ were calculated using the total population of Japan and the population ratios of each prefecture as of January 2021. We used the fmincon programming solver in MATLAB and selected the interior-point algorithm, for which the parameters of the termination tolerances on first-order optimality, function value, and input matrices were set to $1e-4$. The function optimized in fmincon was estimated for each window with a length of $3$ using the $\ell_2$-norm. We used mgcv in R \cite{mgcv2011,mgcv2016} to interpret the control parameters (section \ref{interpret}).

\section{Results and discussion}
\subsection{Evaluating predictive power of hidden controller framework} 
\label{eval-pred-power}
\begin{figure}[htbp]
\begin{tabular}{cc}
\begin{minipage}[t]{0.45\hsize}
\centering
\includegraphics[keepaspectratio, scale=0.4]{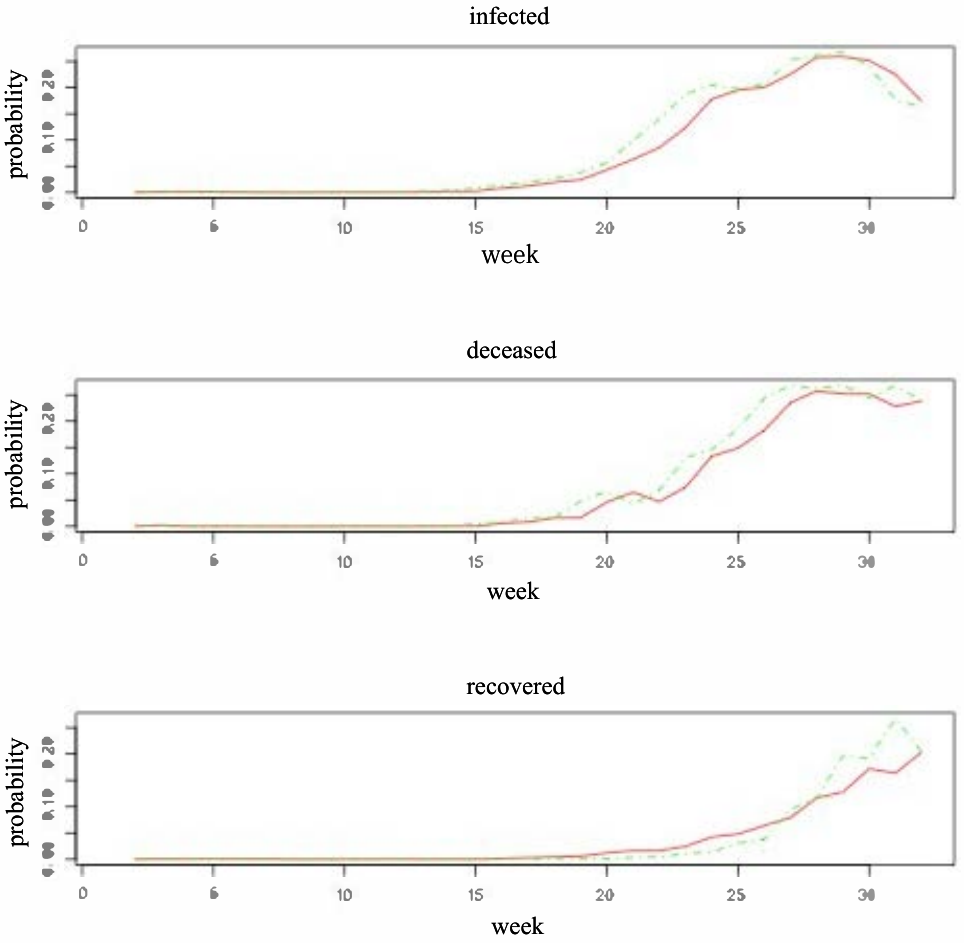}
\subcaption{\centering Australia}
\label{aus}
\end{minipage} &
\begin{minipage}[t]{0.45\hsize}
\centering
\includegraphics[keepaspectratio, scale=0.4]{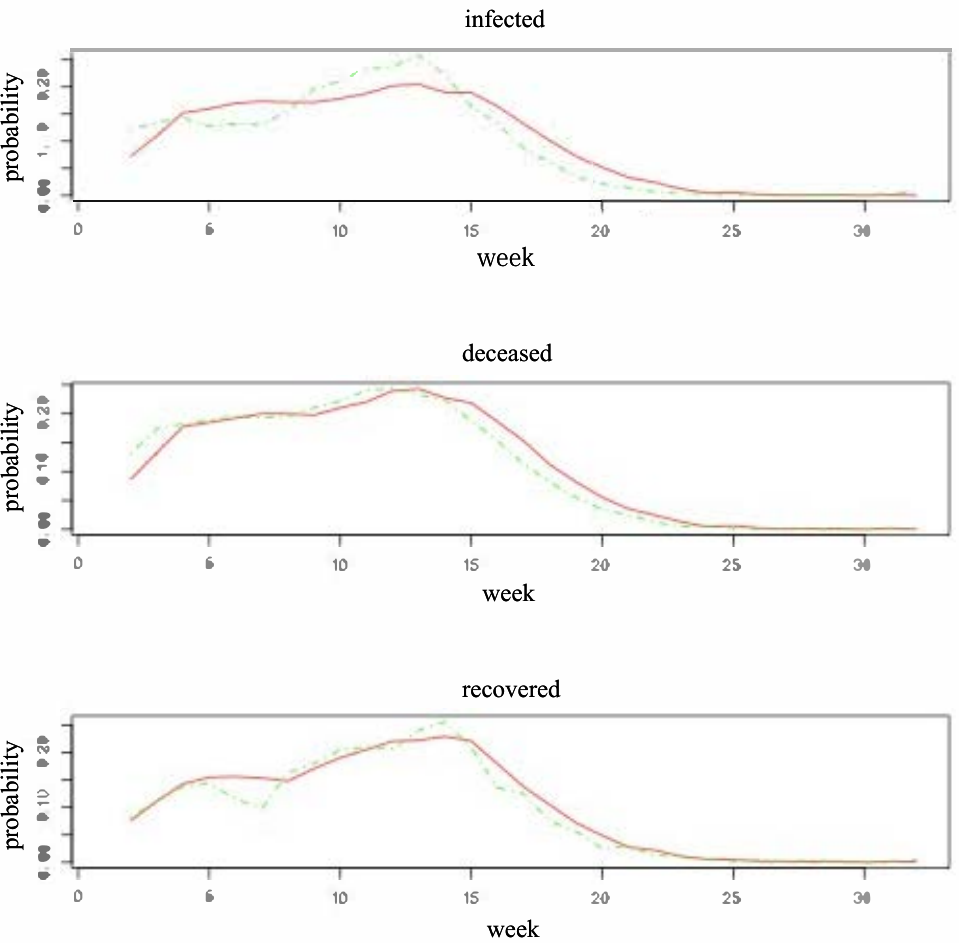}
\subcaption{\centering Colombia}
\label{col}
\end{minipage} \\\\

\begin{minipage}[t]{0.45\hsize}
\centering
\includegraphics[keepaspectratio, scale=0.4]{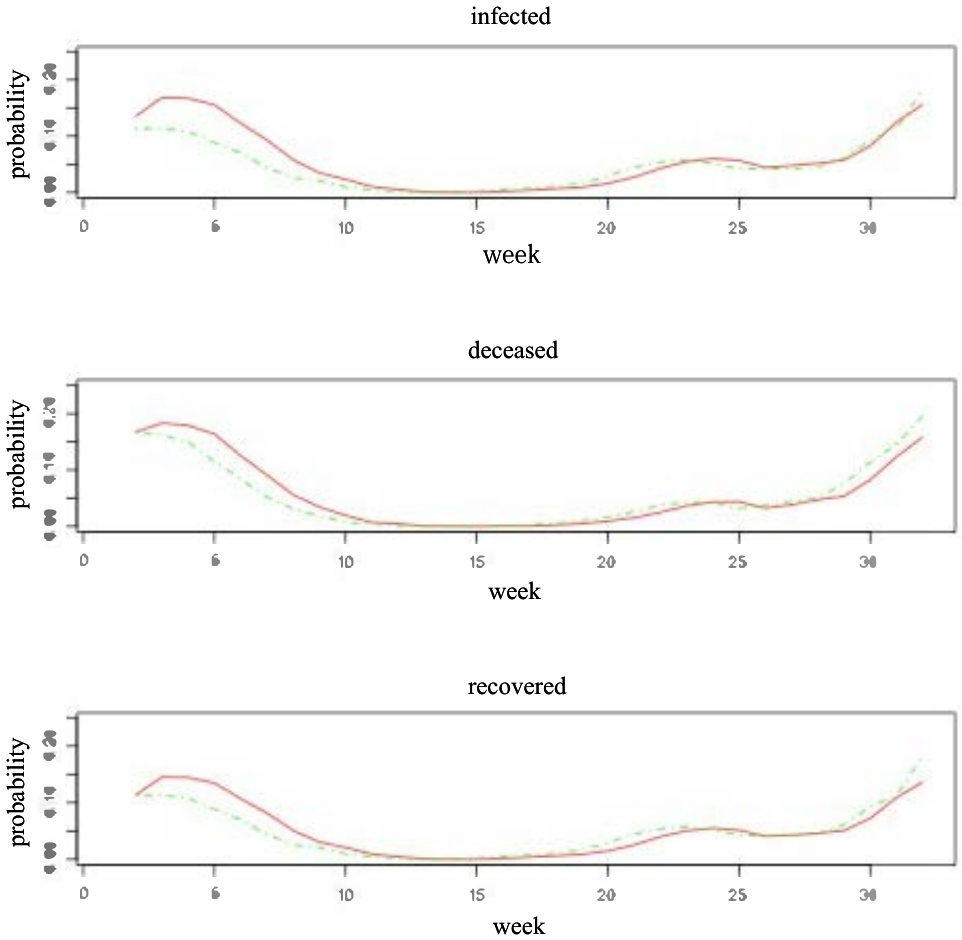}
\subcaption{\centering Germany}
\label{deu}
\end{minipage} &
\begin{minipage}[t]{0.45\hsize}
\centering
\includegraphics[keepaspectratio, scale=0.4]{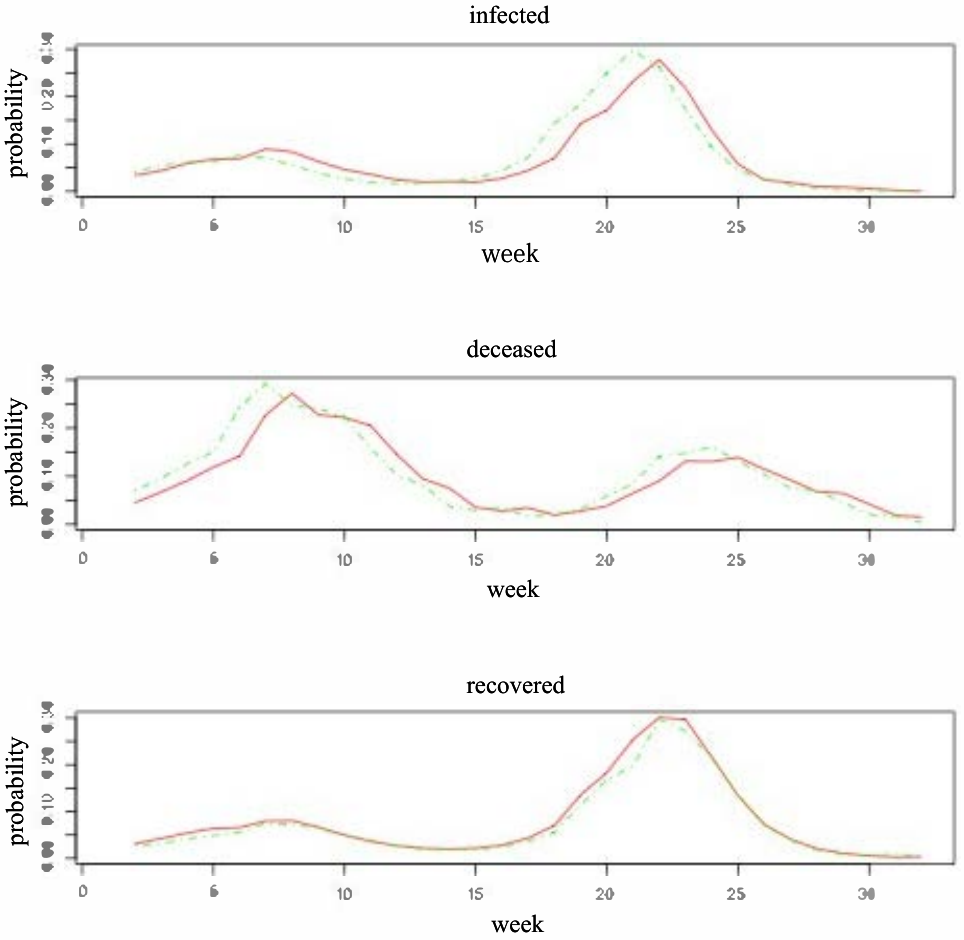}
\subcaption{\centering Japan}
\label{jpn}
\end{minipage} 
\end{tabular}
\caption{Comparison between predictions made by our hidden controller framework (green dashed lines) and observed data (red solid lines) for (a) Australia, (b) Colombia, (c) Germany, and (d) Japan.} 
\label{pred-obse}
\end{figure}

Figure \ref{pred-obse} shows comparisons between our hidden controller framework's prediction ($x^{\ast}_k$) and the observed $x_k$ for Australia, Colombia, Germany, and Japan based on the data for nine countries (data set (ii)). The predictions are reasonably accurate, suggesting that model parameters $A, B$ were well-estimated. Similar results were observed for the other five countries, as detailed in Appendix \ref{other5}. These results underscore the predictive capacity of our hidden controller framework. Furthermore, control feedback $u_k$ can be effectively used in subsequent data analyses.

\subsection{Predictor covariates aimed at halving the number of infected and deceased individuals}
\begin{figure}
\begin{tabular}{cc}
\begin{minipage}[t]{0.5\hsize}
\centering
\includegraphics[keepaspectratio, scale=0.195]{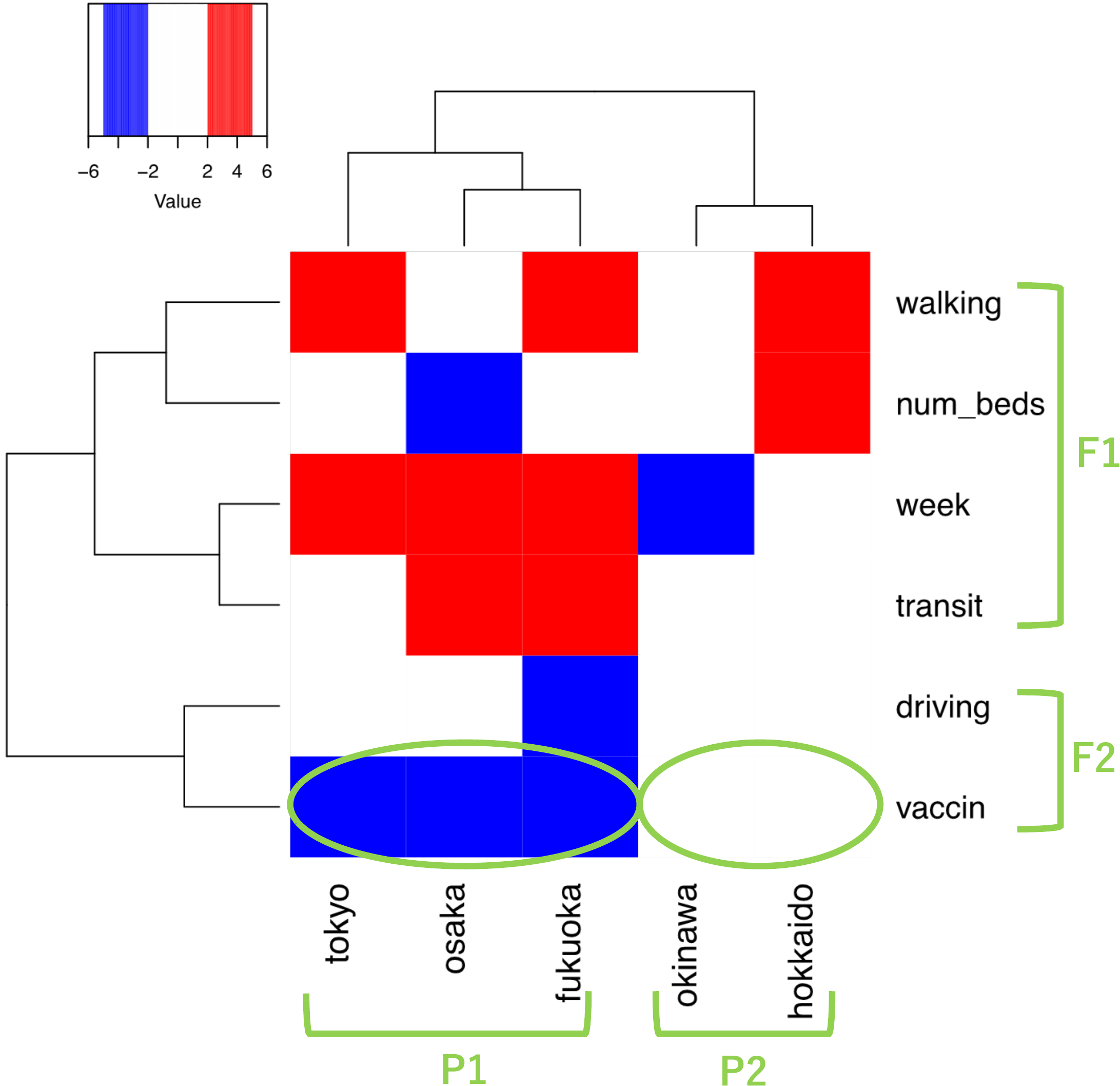}
\subcaption{\centering Infected}
\label{infected-rate}
\end{minipage} &
\begin{minipage}[t]{0.5\hsize}
\centering
\includegraphics[keepaspectratio, scale=0.195]{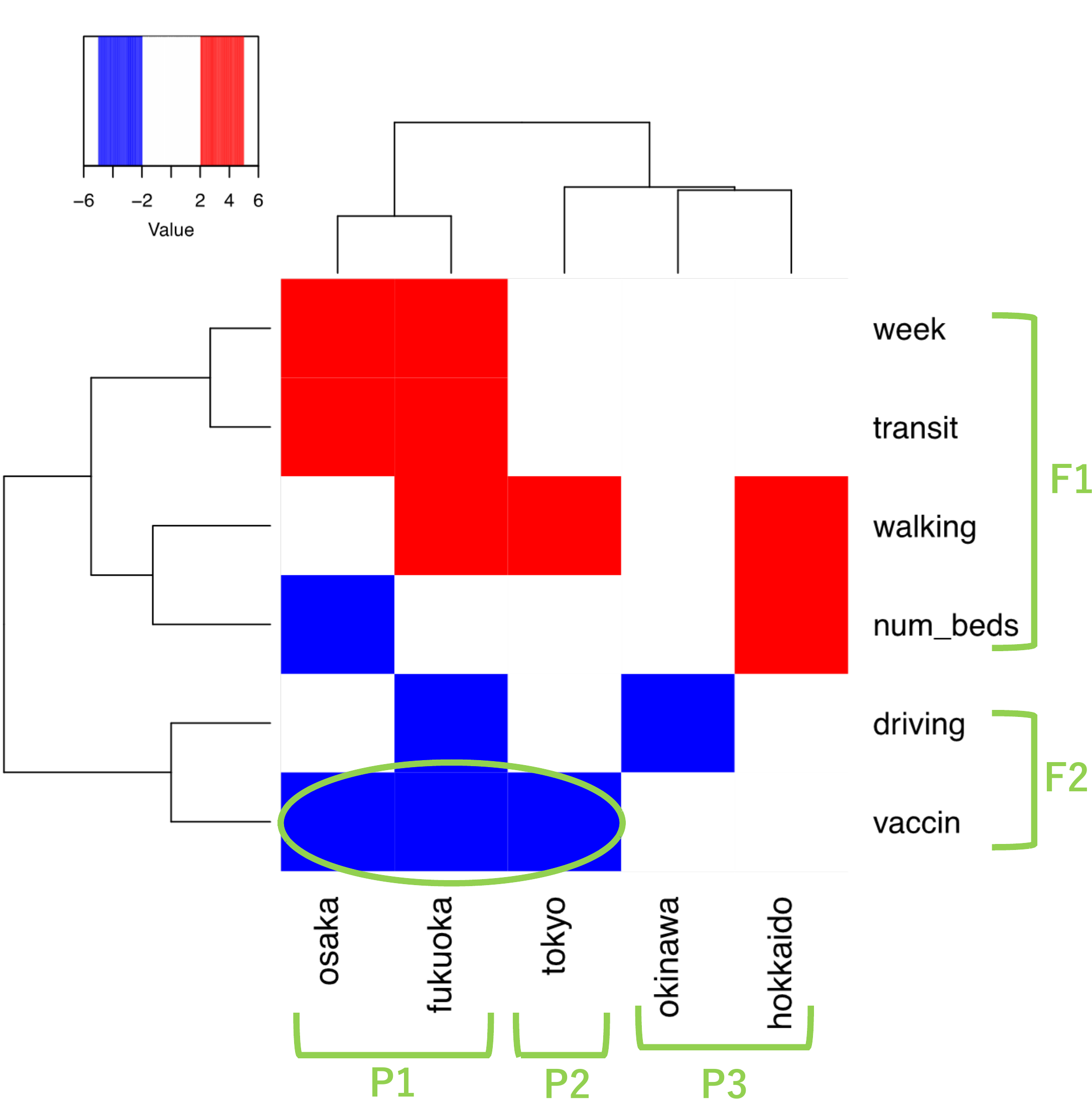}
\subcaption{\centering Deceased}
\label{death-rate}
\end{minipage} 
\end{tabular}
\caption{Heatmaps depicting z-scores of predictor covariates associated with numbers of (a) infected and (b) deceased individuals, with a goal of reducing the counts in both categories by half.}
\label{rate}
\end{figure}

To analyze which predictor covariates effectively reduce the number of infected and deceased individuals, we applied our hidden controller framework (section \ref{Strategy}) and a multivariate GAM (section \ref{interpret}) to a dataset in which the number of infected and deceased individuals was halved (dataset (i)). Figure \ref{rate} shows heatmaps depicting the z-scores of the predictor covariates associated with the numbers of (a) infected and (b) deceased individuals. Since z-scores approximately follow a Student's t-distribution and transition to a normal distribution when the degree of freedom becomes large, we calculated the z-scores at a significance level of 0.05, assuming a normal distribution. We color-coded the heatmaps blue for smaller z-scores, representing positive effects in decreasing the number of infected and deceased individuals, and red for larger z-scores, representing negative effects that increase these numbers.

We roughly categorized the predictor covariates into two groups, F1 and F2. F1 includes "week," "transit," "walking," and "num\_beds," which generally had negative effects. In contrast, the predictor covariates in F2 ("driving" and "vaccination") had positive effects. These results suggest that, in Japan, traveling by automobile and getting vaccinated effectively reduced the number of infected and deceased individuals.

We can classify the prefectures into two groups: the major metropolises of Tokyo, Osaka, and Fukuoka, and the tourist cities in Okinawa and Hokkaido. The data suggest that vaccination in metropolitan areas is highly effective.

\begin{figure}
\begin{tabular}{cc}
\begin{minipage}[t]{0.5\hsize}
\centering
\includegraphics[keepaspectratio, scale=0.195]{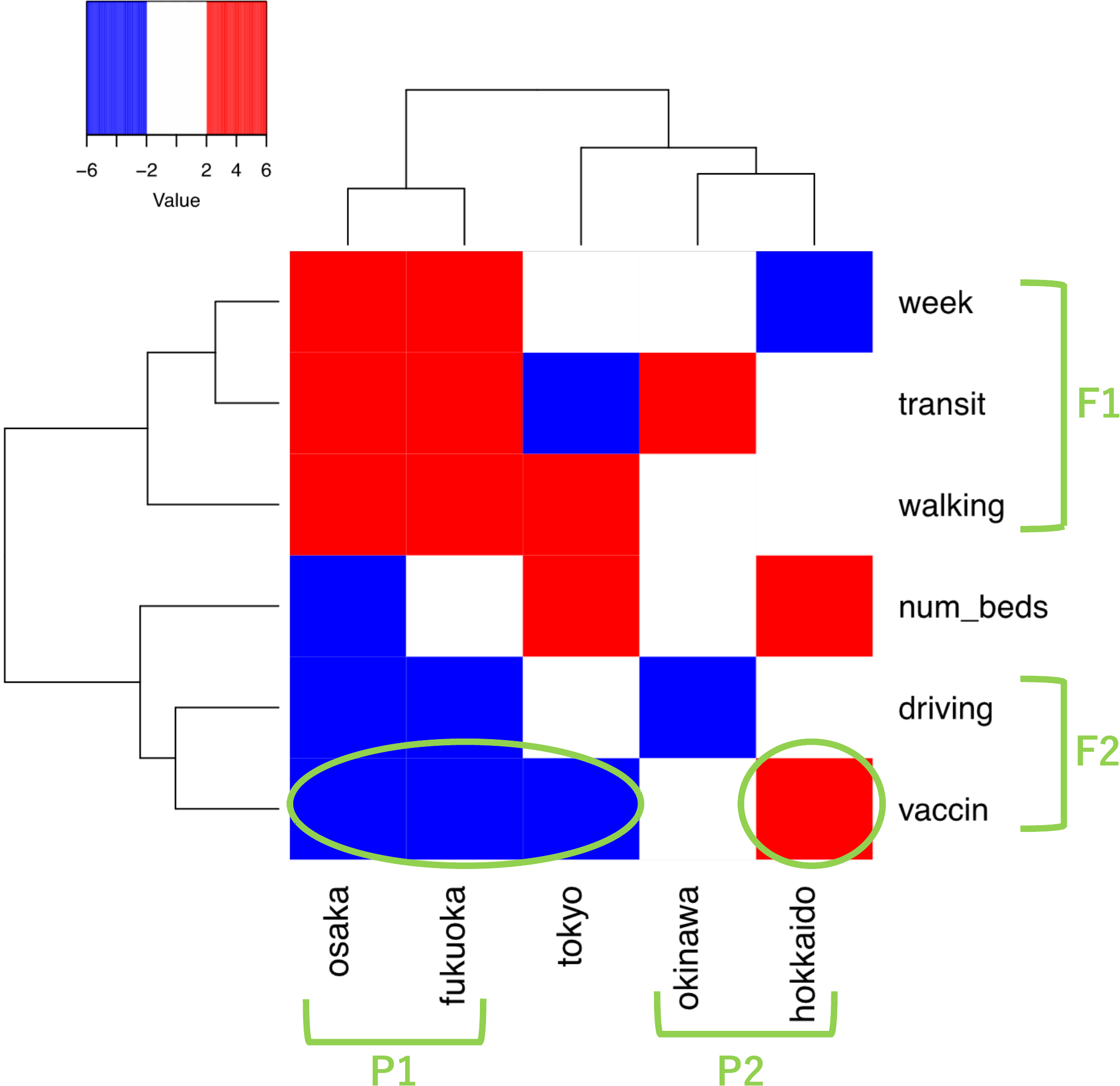}
\subcaption{\centering Infected}
\label{infected-diff}
\end{minipage} &
\begin{minipage}[t]{0.5\hsize}
\centering
\includegraphics[keepaspectratio, scale=0.195]{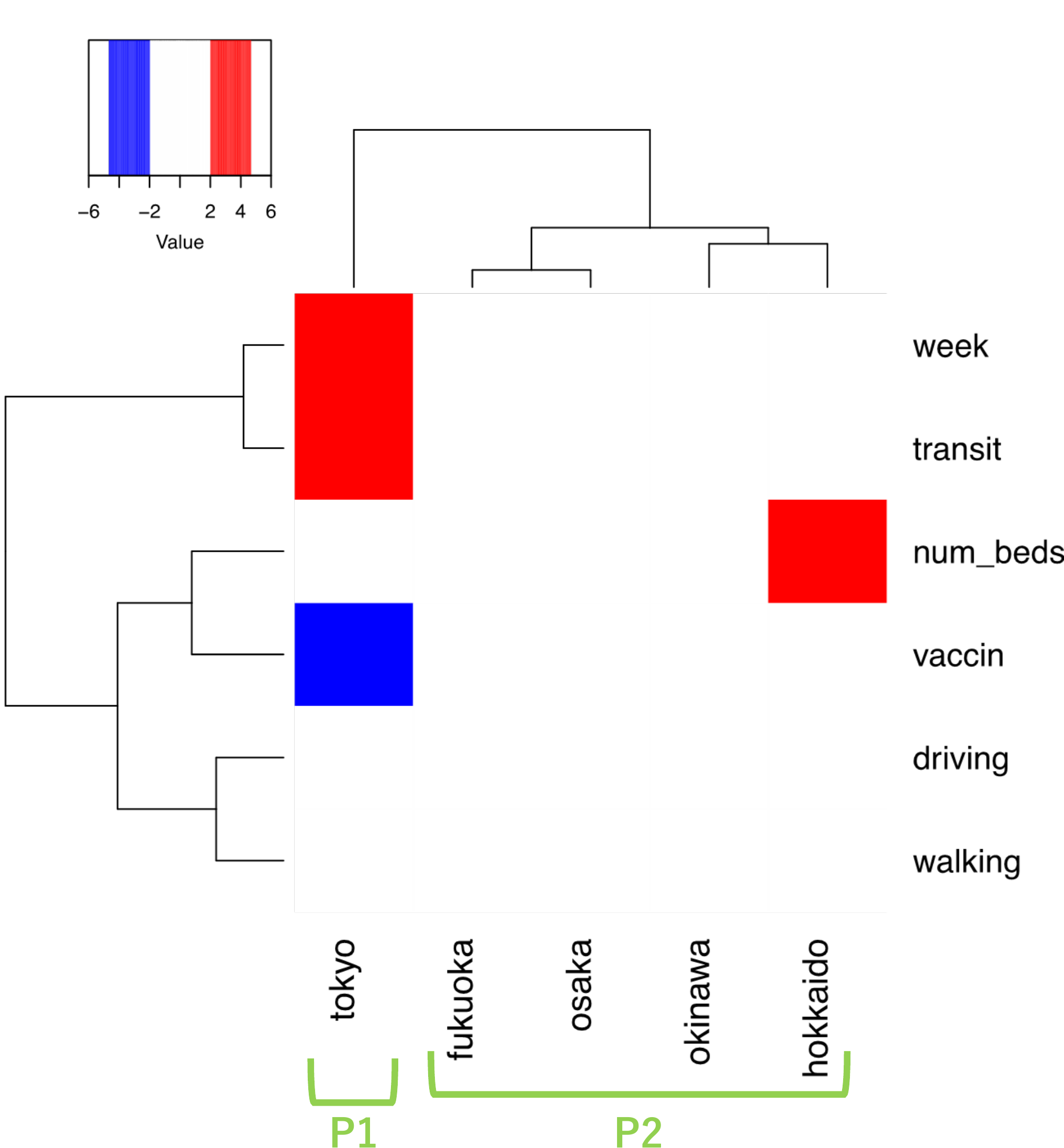}
\subcaption{\centering Deceased}
\label{death-diff}
\end{minipage} 
\end{tabular}
\caption{Heatmaps illustrating difference in z-scores for predictor covariates associated with numbers of (a) infected and (b) deceased individuals, aimed at reducing each category by half compared with the observed numbers. } 
\label{diff}
\end{figure}

To further investigate the positive and negative effects on reducing the number of infected and deceased individuals, we analyzed the difference in z-scores for predictor covariates associated with both categories, with the goal of halving each compared with the observed numbers. Figure \ref{diff} presents heatmaps illustrating these differences in z-scores. For the infected category, vaccination was highly effective in the metropolises of Tokyo, Osaka, and Fukuoka. For the deceased category, vaccination was particularly effective in Tokyo.

\subsection{Differences in predictor covariates across countries} 
\label{sec:country}
\begin{figure}
\centering
\includegraphics[width =0.65 \textwidth]{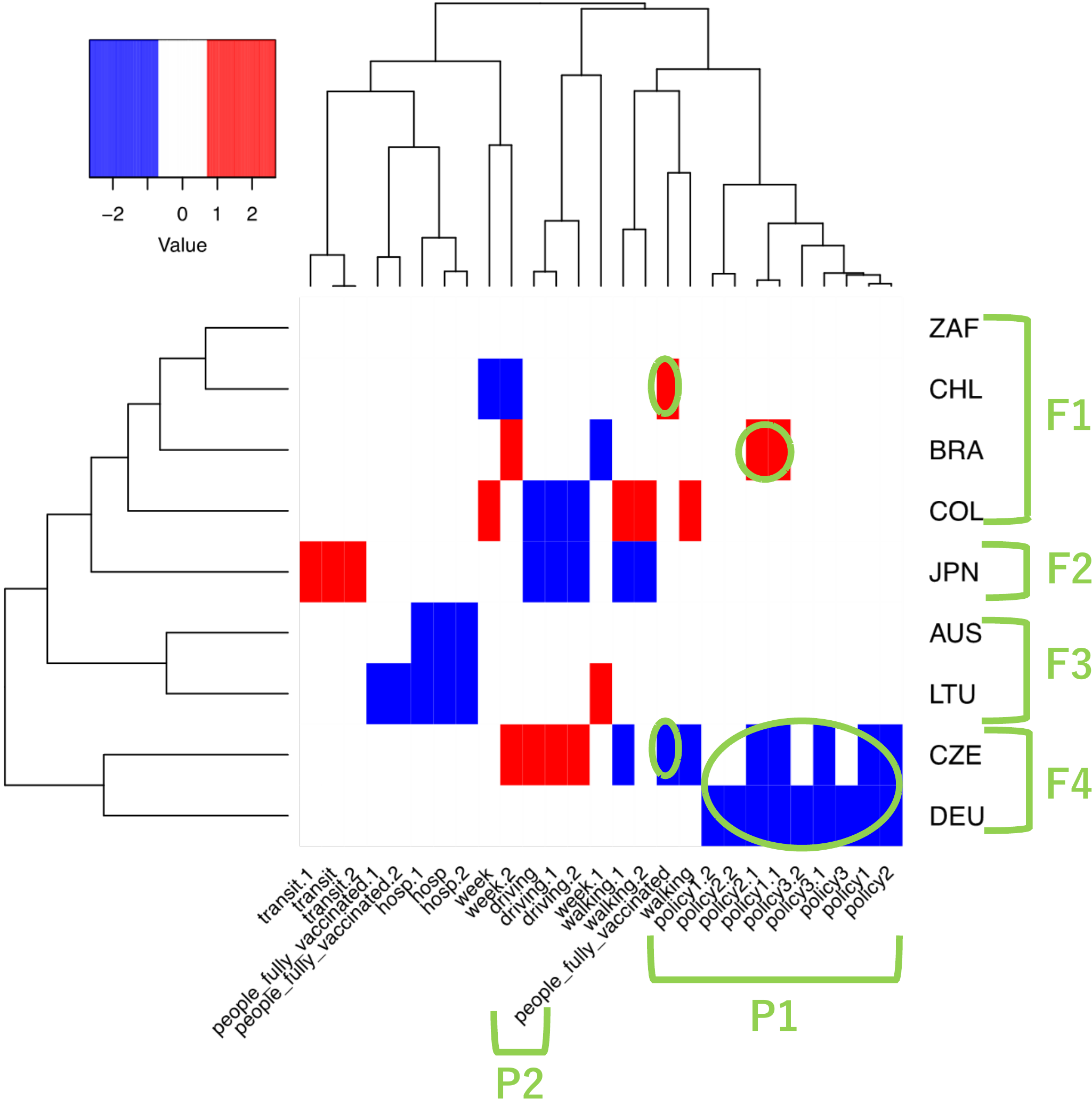}
\caption{Heatmap of z-scores for predictor covariates (cov) associated with number of (a) infected (\{cov\}.1), (b) deceased (\{cov\}.2), and (c) recovered individuals (\{cov\}.3) for nine countries.}
\label{fig:country-cluster}
\end{figure}

Figure \ref{fig:country-cluster} presents a heatmap of z-scores for predictor covariates associated with the number of infected, deceased, and recovered individuals across the nine countries. As an illustration of the classification of characteristics based on control (section \ref{interpret}), the countries can be broadly categorized into four clusters: F1 includes South Africa, Chile, Brazil, and Colombia; F2 is comprised solely of Japan; F3 encompasses Australia and Lithuania; and F4 contains the Czech Republic and Germany. For group F4, three policy measures in Table \ref{policy} (P1 as indicated in the figure) and vaccination (P2) seem to effectively control the number of infected, deceased, and recovered individuals. However, for countries in group F1, the opposite effects were observed.

\subsection{Clustering of countries on basis of policy measures}
\begin{figure}
\centering
\includegraphics[width =0.6 \textwidth]{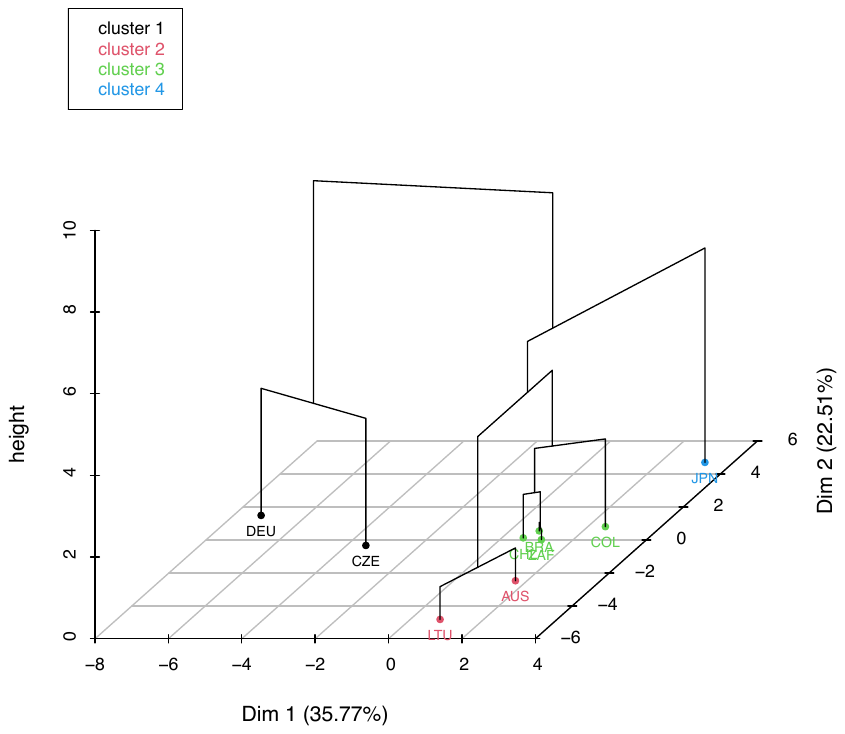}
\caption{Dendrogram illustrating clustering of countries on basis of policy measures.}
\label{fig:dendog}
\end{figure}

Figure \ref{fig:dendog} illustrates the clustering of countries into four groups on the basis of the three categories of policy measures outlined in Table \ref{policy}. These groups align with the ones determined by the z-scores in the previous section. This underscores the critical role of policy measures in controlling the number of infected, deceased, and recovered individuals.

\section{Conclusion}
We have presented a methodology that enables the correlation of various population characteristics with the evolution of a dynamically controlled system. We assume that the observed dynamical system is the result of optimal control of a Kalman-type model, with no direct access to the various stages of this control. The choice of a Kalman-type model was made for simplicity, but this concept can be readily generalized to other models. We assume knowledge of only the model's structure, not its parameters. Our approach, therefore, is to estimate what the controller has done to arrive at the observed data; the estimates of the control parameters are made a posteriori.

A unique aspect of this work is that the evolutionary objects of the dynamical system are probability distributions. We explore how these distributions have evolved under different hidden controller settings. We have provided an algorithm for estimating the various parameters that control the evolution of these distributions, which are based solely on a posteriori observations.

Once the control parameters were obtained, we used a multivariate generalized additive model (GAM) to explore potential relationships between the part that controls the system's evolution and the various characteristics of the studied population. We applied this methodology to a COVID-19 database for five prefectures in Japan, considering various explanatory variables. This approach was generalized to nine countries in order to explore the effects of the various policies they follow.

From the GAM, we extracted indicators to quantify the degrees of association between the different explanatory variables. These new indicators enable the analysis and visualization of groups with similar and different evolutionary profiles. The aim is to explain the reasons for these differences. This work provides a novel perspective on understanding and managing complex evolutionary processes.

\section*{Acknowledgments}
This research was supported by a "Strategic Research Projects" grant from ROIS (Research Organization of Information and Systems), Japan.

\appendix
\section{Additional results for predictive power of hidden controller framework} 
\label{other5}
\begin{figure}[htbp]
\begin{tabular}{cc}
\begin{minipage}[t]{0.45\hsize}
\centering
\includegraphics[keepaspectratio, scale=0.32]{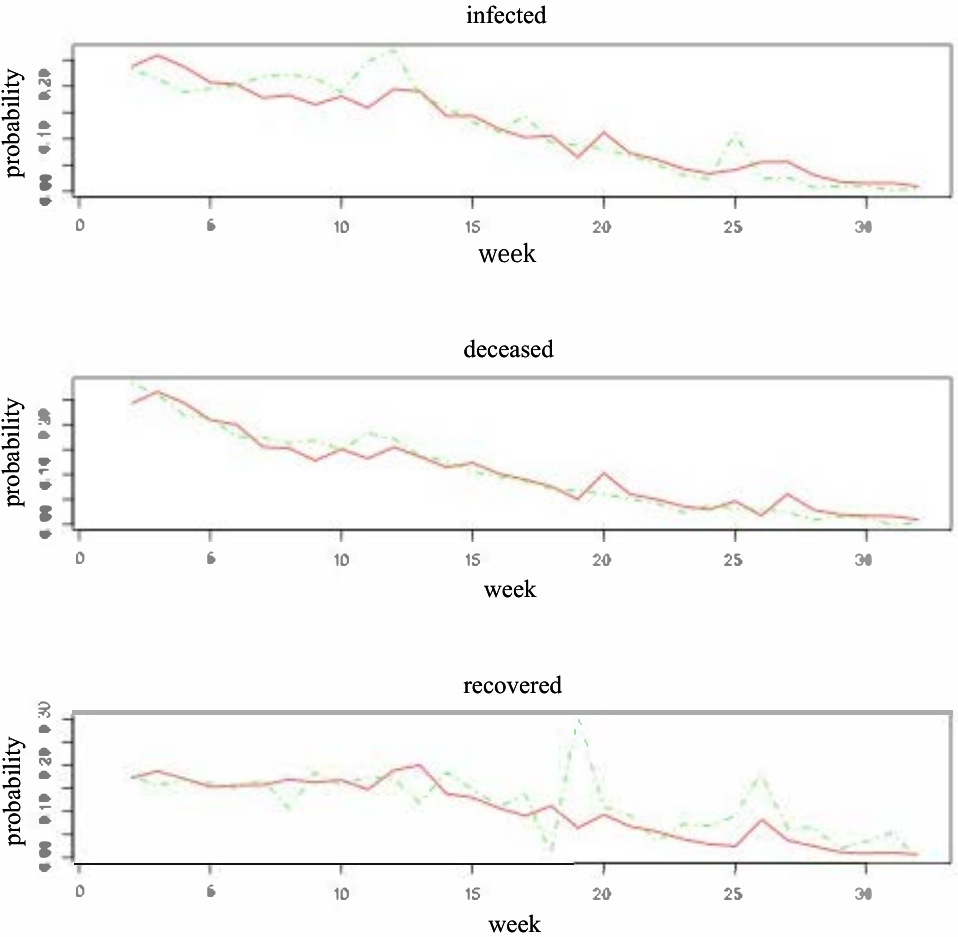}
\subcaption{\centering Brazil}
\label{bra}
\end{minipage} &
\begin{minipage}[t]{0.45\hsize}
\centering
\includegraphics[keepaspectratio, scale=0.32]{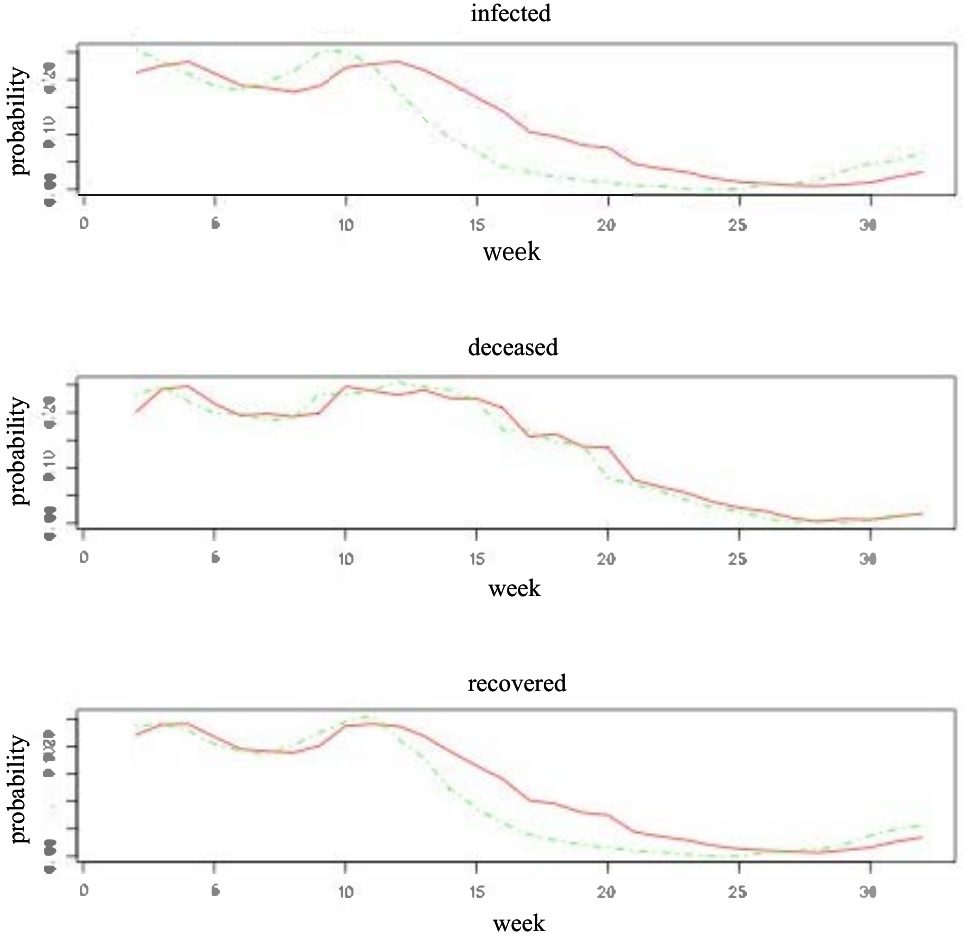}
\subcaption{\centering Chile}
\label{chl}
\end{minipage} \\\\
\begin{minipage}[t]{0.45\hsize}
\centering
\includegraphics[keepaspectratio, scale=0.32]{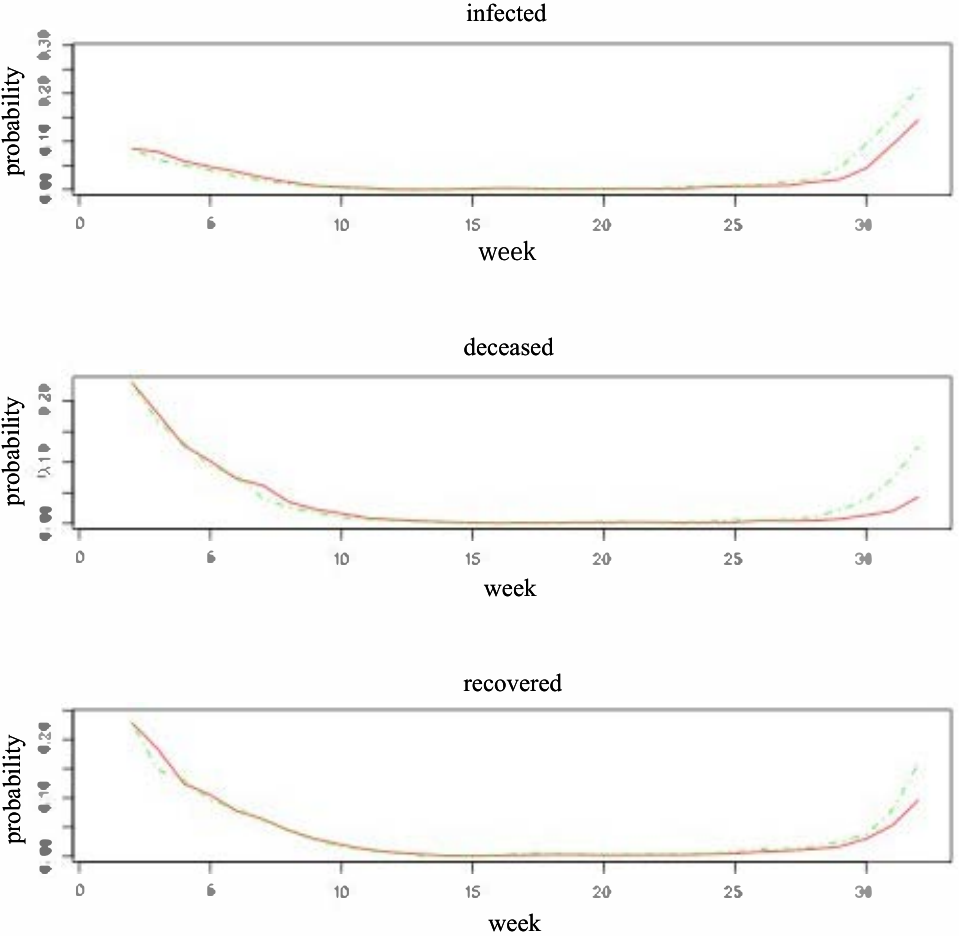}
\subcaption{\centering Czech Republic}
\label{cze}
\end{minipage} &
\begin{minipage}[t]{0.45\hsize}
\centering
\includegraphics[keepaspectratio, scale=0.32]{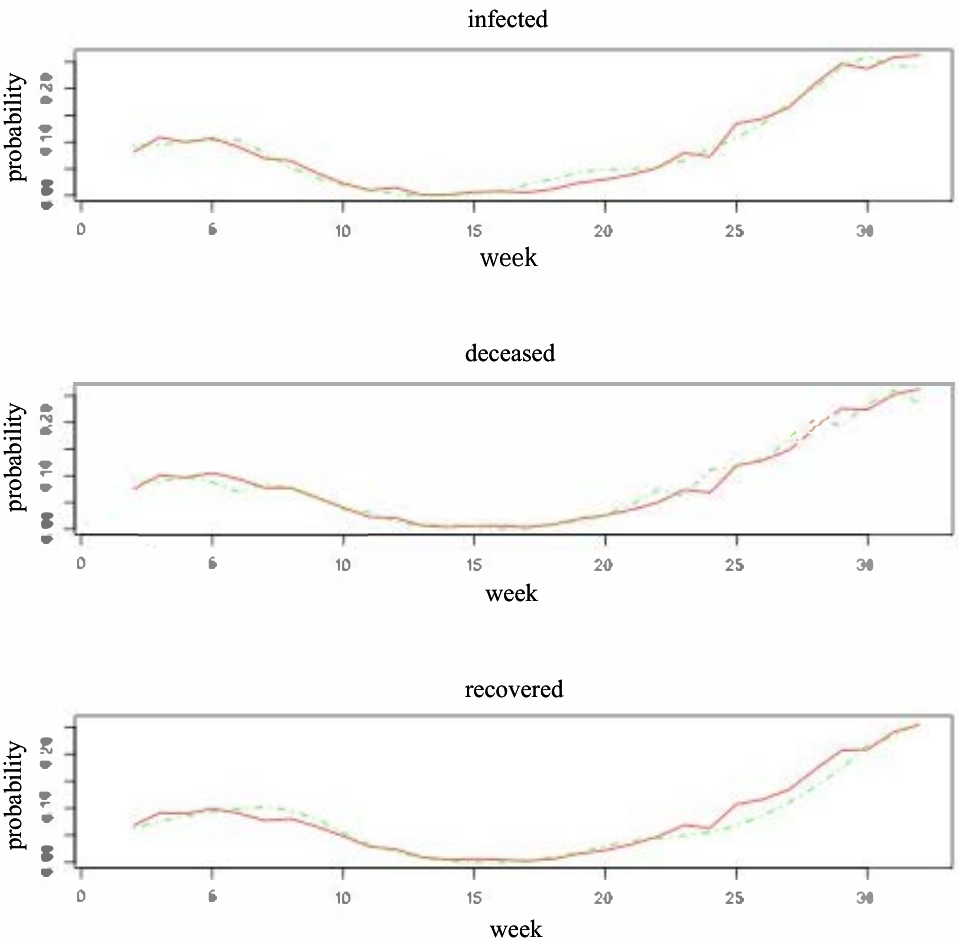}
\subcaption{\centering Lithuania}
\label{ltu}
\end{minipage} \\\\
\begin{minipage}[t]{0.45\hsize}
\centering
\includegraphics[keepaspectratio, scale=0.32]{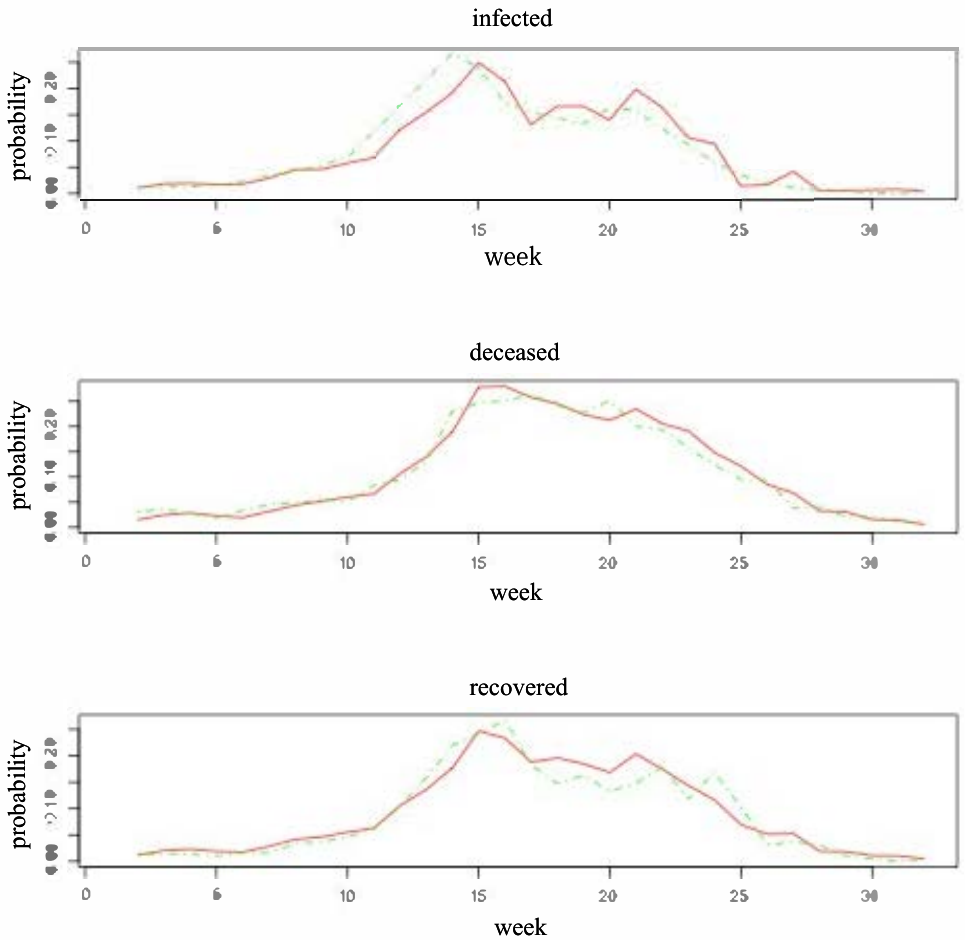}
\subcaption{\centering South Africa}
\label{zaf}
\end{minipage} & 
\end{tabular}
\caption{Comparison between the predictions made by our hidden controller framework (green dashed lines) and observed data (red solid lines) for remaining five countries.} 
\label{pred-obse-other}
\end{figure}

Figure \ref{pred-obse-other} presents a comparison between the predictions made by our hidden controller framework and the observed data for five countries. Consistent with the results in section \ref{eval-pred-power}, our hidden controller framework performed well.

\bibliographystyle{elsarticle-num-names} 
\bibliography{references.bib}

\begin{thebibliography}{25}
\expandafter\ifx\csname natexlab\endcsname\relax\def\natexlab#1{#1}\fi
\providecommand{\url}[1]{\texttt{#1}}
\providecommand{\href}[2]{#2}
\providecommand{\path}[1]{#1}
\providecommand{\DOIprefix}{doi:}
\providecommand{\ArXivprefix}{arXiv:}
\providecommand{\URLprefix}{URL: }
\providecommand{\Pubmedprefix}{pmid:}
\providecommand{\doi}[1]{\href{http://dx.doi.org/#1}{\path{#1}}}
\providecommand{\Pubmed}[1]{\href{pmid:#1}{\path{#1}}}
\providecommand{\bibinfo}[2]{#2}
\ifx\xfnm\relax \def\xfnm[#1]{\unskip,\space#1}\fi
%Type = Article
\bibitem[{Gr{\"u}ne et~al.(2020)Gr{\"u}ne, Schaller, and Schiela}]{grune_exponential_2020}
\bibinfo{author}{L.~Gr{\"u}ne}, \bibinfo{author}{M.~Schaller}, \bibinfo{author}{A.~Schiela},
\newblock \bibinfo{title}{Exponential sensitivity and turnpike analysis for linear quadratic optimal control of general evolution equations},
\newblock \bibinfo{journal}{Journal of Differential Equations} \bibinfo{volume}{268} (\bibinfo{year}{2020}) \bibinfo{pages}{7311--7341}.
%Type = Article
\bibitem[{Bussell et~al.(2019)Bussell, Dangerfield, Gilligan, and Cunniffe}]{bussell_applying_2019}
\bibinfo{author}{E.~H. Bussell}, \bibinfo{author}{C.~E. Dangerfield}, \bibinfo{author}{C.~A. Gilligan}, \bibinfo{author}{N.~J. Cunniffe},
\newblock \bibinfo{title}{Applying optimal control theory to complex epidemiological models to inform real-world disease management},
\newblock \bibinfo{journal}{Philosophical Transactions of the Royal Society B} \bibinfo{volume}{374} (\bibinfo{year}{2019}) \bibinfo{pages}{20180284}.
%Type = Article
\bibitem[{Ros et~al.(2004)Ros, Sorensen, Waagepetersen, Dupont-Nivet, SanCristobal, Bonnet, and Mallard}]{ros_evidence_2004}
\bibinfo{author}{M.~Ros}, \bibinfo{author}{D.~Sorensen}, \bibinfo{author}{R.~Waagepetersen}, \bibinfo{author}{M.~Dupont-Nivet}, \bibinfo{author}{M.~SanCristobal}, \bibinfo{author}{J.-C. Bonnet}, \bibinfo{author}{J.~Mallard},
\newblock \bibinfo{title}{Evidence for genetic control of adult weight plasticity in the snail helix aspersa},
\newblock \bibinfo{journal}{Genetics} \bibinfo{volume}{168} (\bibinfo{year}{2004}) \bibinfo{pages}{2089--2097}.
%Type = Article
\bibitem[{Shen et~al.(2018)Shen, Liu, Yao, Wu, and Yang}]{shen_development_2019}
\bibinfo{author}{L.~Shen}, \bibinfo{author}{R.~Liu}, \bibinfo{author}{Z.~Yao}, \bibinfo{author}{W.~Wu}, \bibinfo{author}{H.~Yang},
\newblock \bibinfo{title}{Development of dynamic platoon dispersion models for predictive traffic signal control},
\newblock \bibinfo{journal}{IEEE Transactions on Intelligent Transportation Systems} \bibinfo{volume}{20} (\bibinfo{year}{2018}) \bibinfo{pages}{431--440}.
%Type = Incollection
\bibitem[{Oosterhoff and van Zwet(2012)}]{HellingerTV}
\bibinfo{author}{J.~Oosterhoff}, \bibinfo{author}{W.~R. van Zwet},
\newblock \bibinfo{title}{A note on contiguity and hellinger distance},
\newblock in: \bibinfo{booktitle}{Selected Works of Willem van Zwet}, \bibinfo{publisher}{Springer}, \bibinfo{year}{2012}, pp. \bibinfo{pages}{63--72}.
%Type = Article
\bibitem[{P{\'e}ni et~al.(2020)P{\'e}ni, Csutak, Szederk{\'e}nyi, and R{\"o}st}]{peni2020nonlinear}
\bibinfo{author}{T.~P{\'e}ni}, \bibinfo{author}{B.~Csutak}, \bibinfo{author}{G.~Szederk{\'e}nyi}, \bibinfo{author}{G.~R{\"o}st},
\newblock \bibinfo{title}{Nonlinear model predictive control with logic constraints for covid-19 management},
\newblock \bibinfo{journal}{Nonlinear Dynamics} \bibinfo{volume}{102} (\bibinfo{year}{2020}) \bibinfo{pages}{1965--1986}.
%Type = Article
\bibitem[{P{\'e}ni and Szederk{\'e}nyi(2021)}]{peni2021convex}
\bibinfo{author}{T.~P{\'e}ni}, \bibinfo{author}{G.~Szederk{\'e}nyi},
\newblock \bibinfo{title}{Convex output feedback model predictive control for mitigation of covid-19 pandemic},
\newblock \bibinfo{journal}{Annual Reviews in Control} \bibinfo{volume}{52} (\bibinfo{year}{2021}) \bibinfo{pages}{543--553}.
%Type = Article
\bibitem[{K{\"o}hler et~al.(2021)K{\"o}hler, Schwenkel, Koch, Berberich, Pauli, and Allg{\"o}wer}]{kohler2021robust}
\bibinfo{author}{J.~K{\"o}hler}, \bibinfo{author}{L.~Schwenkel}, \bibinfo{author}{A.~Koch}, \bibinfo{author}{J.~Berberich}, \bibinfo{author}{P.~Pauli}, \bibinfo{author}{F.~Allg{\"o}wer},
\newblock \bibinfo{title}{Robust and optimal predictive control of the covid-19 outbreak},
\newblock \bibinfo{journal}{Annual Reviews in Control} \bibinfo{volume}{51} (\bibinfo{year}{2021}) \bibinfo{pages}{525--539}.
%Type = Article
\bibitem[{Carli et~al.(2020)Carli, Cavone, Epicoco, Scarabaggio, and Dotoli}]{carli2020model}
\bibinfo{author}{R.~Carli}, \bibinfo{author}{G.~Cavone}, \bibinfo{author}{N.~Epicoco}, \bibinfo{author}{P.~Scarabaggio}, \bibinfo{author}{M.~Dotoli},
\newblock \bibinfo{title}{Model predictive control to mitigate the covid-19 outbreak in a multi-region scenario},
\newblock \bibinfo{journal}{Annual Reviews in Control} \bibinfo{volume}{50} (\bibinfo{year}{2020}) \bibinfo{pages}{373--393}.
%Type = Article
\bibitem[{Morato et~al.(2020)Morato, Bastos, Cajueiro, and Normey-Rico}]{morato2020optimal}
\bibinfo{author}{M.~M. Morato}, \bibinfo{author}{S.~B. Bastos}, \bibinfo{author}{D.~O. Cajueiro}, \bibinfo{author}{J.~E. Normey-Rico},
\newblock \bibinfo{title}{An optimal predictive control strategy for covid-19 (sars-cov-2) social distancing policies in brazil},
\newblock \bibinfo{journal}{Annual reviews in control} \bibinfo{volume}{50} (\bibinfo{year}{2020}) \bibinfo{pages}{417--431}.
%Type = Book
\bibitem[{Brogan(1991)}]{MCT}
\bibinfo{author}{W.~L. Brogan}, \bibinfo{title}{Modern control theory}, \bibinfo{publisher}{Pearson education india}, \bibinfo{year}{1991}.
%Type = Book
\bibitem[{Bryson and Ho(2018)}]{OAOC}
\bibinfo{author}{A.~E. Bryson}, \bibinfo{author}{Y.-C. Ho}, \bibinfo{title}{Applied optimal control: optimization, estimation, and control}, \bibinfo{publisher}{Routledge}, \bibinfo{year}{2018}.
%Type = Book
\bibitem[{Chen(1984)}]{LSTD}
\bibinfo{author}{C.-T. Chen}, \bibinfo{title}{Linear system theory and design}, \bibinfo{publisher}{Saunders college publishing}, \bibinfo{year}{1984}.
%Type = Book
\bibitem[{Fadali and Visioli(2012)}]{DCE}
\bibinfo{author}{M.~S. Fadali}, \bibinfo{author}{A.~Visioli}, \bibinfo{title}{Digital control engineering: analysis and design}, \bibinfo{publisher}{Academic Press}, \bibinfo{year}{2012}.
%Type = Book
\bibitem[{Bittanti et~al.(2012)Bittanti, Laub, and Willems}]{Ricatti}
\bibinfo{author}{S.~Bittanti}, \bibinfo{author}{A.~J. Laub}, \bibinfo{author}{J.~C. Willems}, \bibinfo{title}{The Riccati Equation}, \bibinfo{publisher}{Springer Science \& Business Media}, \bibinfo{year}{2012}.
%Type = Article
\bibitem[{de~Gouv{\^e}a and Odloak(1997)}]{SQP}
\bibinfo{author}{M.~de~Gouv{\^e}a}, \bibinfo{author}{D.~Odloak},
\newblock \bibinfo{title}{Dealing with inconsistent quadratic programs in a sqp based algorithm},
\newblock \bibinfo{journal}{Brazilian Journal of Chemical Engineering} \bibinfo{volume}{14} (\bibinfo{year}{1997}) \bibinfo{pages}{63--80}.
%Type = Article
\bibitem[{Wright et~al.(1999)Wright, Nocedal et~al.}]{NO}
\bibinfo{author}{S.~Wright}, \bibinfo{author}{J.~Nocedal}, et~al.,
\newblock \bibinfo{title}{Numerical optimization},
\newblock \bibinfo{journal}{Springer Science} \bibinfo{volume}{35} (\bibinfo{year}{1999}) \bibinfo{pages}{7}.
%Type = Book
\bibitem[{Fletcher(2013)}]{PMOO}
\bibinfo{author}{R.~Fletcher}, \bibinfo{title}{Practical methods of optimization}, \bibinfo{publisher}{John Wiley \& Sons}, \bibinfo{year}{2013}.
%Type = Inproceedings
\bibitem[{Schittkowski and Zillober(2003)}]{NLPASA}
\bibinfo{author}{K.~Schittkowski}, \bibinfo{author}{C.~Zillober},
\newblock \bibinfo{title}{Nonlinear programming: algorithms, software, and applications},
\newblock in: \bibinfo{booktitle}{IFIP Conference on System Modeling and Optimization}, \bibinfo{organization}{Springer}, \bibinfo{year}{2003}, pp. \bibinfo{pages}{73--107}.
%Type = Article
\bibitem[{Wood(2011)}]{mgcv2011}
\bibinfo{author}{S.~N. Wood},
\newblock \bibinfo{title}{Fast stable restricted maximum likelihood and marginal likelihood estimation of semiparametric generalized linear models},
\newblock \bibinfo{journal}{Journal of the Royal Statistical Society (B)} \bibinfo{volume}{73} (\bibinfo{year}{2011}) \bibinfo{pages}{3--36}.
%Type = Article
\bibitem[{Wood et~al.(2016)Wood, {N.}, {Pya}, and S{"a}fken}]{mgcv2016}
\bibinfo{author}{S.~Wood}, \bibinfo{author}{{N.}}, \bibinfo{author}{{Pya}}, \bibinfo{author}{B.~S{"a}fken},
\newblock \bibinfo{title}{Smoothing parameter and model selection for general smooth models (with discussion)},
\newblock \bibinfo{journal}{Journal of the American Statistical Association} \bibinfo{volume}{111} (\bibinfo{year}{2016}) \bibinfo{pages}{1548--1575}.
%Type = Article
\bibitem[{Guidotti(2022)}]{COVID19data}
\bibinfo{author}{E.~Guidotti},
\newblock \bibinfo{title}{A worldwide epidemiological database for covid-19 at fine-grained spatial resolution},
\newblock \bibinfo{journal}{Scientific Data} \bibinfo{volume}{9} (\bibinfo{year}{2022}) \bibinfo{pages}{112}. \DOIprefix\doi{10.1038/s41597-022-01245-1}.
%Type = Article
\bibitem[{Guidotti and Ardia(2020)}]{COVID19github}
\bibinfo{author}{E.~Guidotti}, \bibinfo{author}{D.~Ardia},
\newblock \bibinfo{title}{Covid-19 data hub},
\newblock \bibinfo{journal}{Journal of Open Source Software} \bibinfo{volume}{5} (\bibinfo{year}{2020}) \bibinfo{pages}{2376}. \DOIprefix\doi{10.21105/joss.02376}.
%Type = Article
\bibitem[{Matsui et~al.(2022)Matsui, Azzaoui, and Murakami}]{matsui_analysis_2022}
\bibinfo{author}{T.~Matsui}, \bibinfo{author}{N.~Azzaoui}, \bibinfo{author}{D.~Murakami},
\newblock \bibinfo{title}{Analysis of {COVID}-19 evolution based on testing closeness of sequential data}  (\bibinfo{year}{2022}). \URLprefix \url{https://doi.org/10.1007/s42081-021-00144-w}. \DOIprefix\doi{10.1007/s42081-021-00144-w}.
%Type = Article
\bibitem[{Hale et~al.(2021)Hale, Angrist, Goldszmidt, Kira, Petherick, Phillips, Webster, Cameron-Blake, Hallas, Majumdar et~al.}]{hale2021global}
\bibinfo{author}{T.~Hale}, \bibinfo{author}{N.~Angrist}, \bibinfo{author}{R.~Goldszmidt}, \bibinfo{author}{B.~Kira}, \bibinfo{author}{A.~Petherick}, \bibinfo{author}{T.~Phillips}, \bibinfo{author}{S.~Webster}, \bibinfo{author}{E.~Cameron-Blake}, \bibinfo{author}{L.~Hallas}, \bibinfo{author}{S.~Majumdar}, et~al.,
\newblock \bibinfo{title}{A global panel database of pandemic policies (oxford covid-19 government response tracker)},
\newblock \bibinfo{journal}{Nature human behaviour} \bibinfo{volume}{5} (\bibinfo{year}{2021}) \bibinfo{pages}{529--538}.

\end{thebibliography}

\end{document}